\documentclass[aps,plb,letterpaper,twocolumn,reprint,preprintnumbers,
floatfix,superscriptaddress,nofootinbib,showpacs]{revtex4-1}
\usepackage{graphicx}
\usepackage{dcolumn}
\usepackage{bm}
\usepackage{epsfig,cancel}
\usepackage{epstopdf}
\usepackage{amsmath}
\usepackage{amssymb}
\usepackage{graphicx,bm}
\usepackage{color}
\usepackage{hyperref}
\usepackage{url}
\usepackage{breakurl}

\newcommand{\be}{\begin{equation}}
\newcommand{\ee}{\end{equation}}
\newcommand{\bea}{\begin{eqnarray}}
\newcommand{\eea}{\end{eqnarray}}
\def\met{\slash{\!\!\!\!E}_{\text{T}}}
\def\pT{p_{\text{T}}}

\usepackage{color}

\begin{document}

\title{The 750GeV diphoton excess: who introduces it?}

\author{Hao Zhang}
\affiliation{Department of Physics, University of California, Santa Barbara, California 93106, USA}

\begin{abstract}
Recently, both ATLAS and CMS collaborations report an excess at 750GeV 
in the diphoton invariant mass spectrum at 13TeV LHC. If it is a
new scalar produced via loop induced gluon-gluon fusion process,
it is important to know what is the particle in the loop. In this work,
we investigate the possibility of determine the fraction of the contribution from 
the standard model top-quark in the loop.
\end{abstract}

\maketitle

\section{Introduction}
\label{sec:intro}
An excess in diphoton invariant mass spectrum is reported by both 
ATLAS and CMS collaborations at 13TeV LHC \cite{ATLAS-CONF-2015-081,
CMS-PAS-EXO-15-004}. Although more data is needed to make 
definite conclusion, it is probably a hint of a new resonance at 750GeV 
and width is 45GeV. In a few weeks, more than a hundred papers
appear online to explain the excess \cite{Jaeckel:2012yz,
Harigaya:2015ezk,Mambrini:2015wyu,
Backovic:2015fnp,Angelescu:2015uiz,Nakai:2015ptz,Knapen:2015dap,Buttazzo:2015txu,
Pilaftsis:2015ycr,Franceschini:2015kwy,DiChiara:2015vdm,McDermott:2015sck,
Ellis:2015oso,Low:2015qep,Bellazzini:2015nxw,Gupta:2015zzs,Petersson:2015mkr,
Molinaro:2015cwg,Costa:2015llh,Dutta:2015wqh,Cao:2015pto,Yamatsu:2015oit,
Matsuzaki:2015che,Kobakhidze:2015ldh,Martinez:2015kmn,Cox:2015ckc,Becirevic:2015fmu,
No:2015bsn,Demidov:2015zqn,Gopalakrishna:2015dkt,Chao:2015ttq,Fichet:2015vvy,
Bian:2015kjt,Chakrabortty:2015hff,Ahmed:2015uqt,Agrawal:2015dbf,Csaki:2015vek,
Falkowski:2015swt,Aloni:2015mxa,Bai:2015nbs,Gabrielli:2015dhk,Benbrik:2015fyz,
Kim:2015ron,Alves:2015jgx,Megias:2015ory,Carpenter:2015ucu,Chao:2015nsm,
Arun:2015ubr,Han:2015cty,Chang:2015bzc,Chakraborty:2015jvs,Ding:2015rxx,
Han:2015dlp,Han:2015qqj,Luo:2015yio,Chang:2015sdy,Bardhan:2015hcr,Feng:2015wil,
Wang:2015kuj,Antipin:2015kgh,Cao:2015twy,Huang:2015evq,Liao:2015tow,Heckman:2015kqk,
Dhuria:2015ufo,Bi:2015uqd,Kim:2015ksf,Berthier:2015vbb,Cho:2015nxy,Cline:2015msi,
Chala:2015cev,Barducci:2015gtd,Boucenna:2015pav,Murphy:2015kag,Hernandez:2015ywg,
Dey:2015bur,Pelaggi:2015knk,deBlas:2015hlv,Belyaev:2015hgo,Dev:2015isx,Huang:2015rkj,
Moretti:2015pbj,Patel:2015ulo,Badziak:2015zez,Chakraborty:2015gyj,Cao:2015xjz,
Altmannshofer:2015xfo,Cvetic:2015vit,Gu:2015lxj,Allanach:2015ixl,Davoudiasl:2015cuo,
Craig:2015lra,Das:2015enc,Cheung:2015cug,Liu:2015yec,Zhang:2015uuo,Casas:2015blx,
Hall:2015xds,Han:2015yjk,Park:2015ysf,Salvio:2015jgu,Chway:2015lzg,Li:2015jwd,Son:2015vfl,
Tang:2015eko,An:2015cgp,Cao:2015apa,Wang:2015omi,Cai:2015hzc,Cao:2015scs,
Kim:2015xyn,Gao:2015igz,Chao:2015nac,Bi:2015lcf,Goertz:2015nkp,Anchordoqui:2015jxc,
Dev:2015vjd,Bizot:2015qqo,Ibanez:2015uok,Chiang:2015tqz,Kang:2015roj,Hamada:2015skp,
Huang:2015svl,Kanemura:2015bli,Kanemura:2015vcb,Low:2015qho,Hernandez:2015hrt,
Jiang:2015oms,Kaneta:2015qpf,Marzola:2015xbh,Ma:2015xmf,Dasgupta:2015pbr,
Jung:2015etr,Potter:2016psi,Palti:2016kew,Nomura:2016fzs,Han:2016bus,Ko:2016lai,
Ghorbani:2016jdq,Danielsson:2016nyy,Chao:2016mtn,Csaki:2016raa,Karozas:2016hcp,
Hernandez:2016rbi,Modak:2016ung,Dutta:2016jqn,Deppisch:2016scs}. A new
scalar which is produced via the gluon-gluon fusion 
(ggF) channel at the LHC is one of the most popular candidate of this 
excess. In a former work, we discuss the possibility
of distinguishing the $q\bar q$ and $b\bar b$ initial state production
channel from the ggF \cite{Gao:2015igz}. 
We showed that we can know whether the ggF process is the dominant 
production mode in the near future. If ggF 
is the dominant production mode, it is important 
to know where is this loop induced effective operator from. 
Is there a significant contribution from exotic colored particle in the loop?

In this work, we try to answer this question. 
If the excess is confirmed by data in the future, we 
suggest the experimentalists 
look for the $t\bar t\gamma\gamma$ signal at the LHC
Run-II, which can be used to measure the top-quark 
contribution in the loop. The reasons are explained in detail
in Sec. \ref{sec:signal}. In Sec. \ref{sec:pheno}, we 
study the LHC phenomenology of this signal. 
A simple simulation for a 100TeV $pp$ collider is also
shown there. Our conclusions are summarized
in Sec. \ref{sec:concl}. 

\section{The signal}
\label{sec:signal}
We limit our discussion on a 750 GeV scalar resonance $\phi$
produced at the LHC via effective operator\footnote{If the 
particles in the loop is from new physics (NP) at cutoff scale $\Lambda$, 
from the point of the effective field theory view, the interaction
should be expanded in according to the order of $1/\Lambda$
but not $1/v$ where $v$ is the scale of the electroweak spontaneously symmetry
breaking (ESSB). Here we absorb the 
cutoff scale $\Lambda$ into $c_\phi$ for formally simplicity. }
\be
\frac{\alpha_sc_\phi}{12\pi v}\phi G^a_{\mu\nu}G^{a,\mu\nu}.
\ee
Such a loop-induced effective operator could be generated 
through a SM top-quark loop or some colored NP
particles. If the operator is from top-quark loop, it is well known 
that the amplitude square could be written as\footnote{
There is an assumption that the Lorentz structure of the
interaction between $\phi$ and the SM top-quark is 
$\phi\bar tt$. The discussion for a pseudo-scalar with 
$\phi\bar t\gamma_5t$ interaction and a generic scalar 
with $\phi\bar te^{i\theta\gamma_5}t$ is similarly. People 
can discuss non-renormalizable interactions between 
$\phi$ and the SM top-quark. However, the existence 
of them means NP effective in the production.}
\bea
\left|\mathcal{M}\left(gg\to \phi\right)\right|^2
&=&\frac{\alpha^2_sc_t^2G_FM_\phi^4}{96\sqrt2\pi^2}
\biggl|\frac{m_t^2}{M_\phi^2}\biggl[2+\left(1-\frac{4m_t^2}
{M_\phi^2}\right)\nonumber\\
&&\times f\left(\frac{m_t^2}
{M_\phi^2}\right)\biggr]\biggr|^2,
\eea
where
\bea
f\left(x\right)\equiv\left\{\begin{array}{cc}2\arcsin^2\left(\frac{1}{2\sqrt x}\right), & x>\frac{1}{4}, \\
-\frac{1}{2}\left[\log\left(\frac{1+\sqrt{1-4x}}{1-\sqrt{1-4x}}-i\pi\right)\right]^2, & 
x<\frac{1}{4}. \end{array}\right.
\eea
Here we introduce $c_t$ to describe the contribution to 
$c_\phi$ from the top-quark loop. We have
\be
c_\phi=c_t+c_{NP},
\ee
where $c_{NP}$ is the 
contribution to $c_\phi$ from the exotic colored particles.
Our aim is investigating the size of $c_{NP}$
and $c_t$. The 
$pp\to \phi\to t\bar t$ is one of the possible channel. However,
there are some disadvantages of this channel. First, the result 
of this channel depends on
\be
\frac{{\text{Br}}\left(\phi\to \gamma\gamma\right)}
{{\text{Br}}\left(\phi\to t\bar t\right)}.
\ee
This ratio is highly model dependent. Even when we fix the production 
mechanism, it still depend on the details in the $\phi$ decay. Such a 
dependence will 
make the conclusion weaker. Second, it has been well known 
that the ``peak'' in the $t\bar t$ invariant mass spectrum from 
this process is suffered by the interference effect with the SM 
top-pair production \cite{Craig:2015jba,Bernreuther:2015fts}. 
This interference effect will smear the ``peak''
and make the discovery of the signal very difficult at the LHC,
especially for the $\phi$ which is heavier than 700GeV. These reasons make 
the $pp\to \phi\to t\bar t$ not be a good channel to investigate the 
contribution of the top-quark in the $\phi$ production. The $pp\to
t\bar t\phi\to t\bar tt\bar t$ channel is helpful to solve the second 
problem \cite{Han:2004zh}. But it still highly depends on the details of the decay of the
resonance. Because of the large SM backgrounds, at least $\sim$1fb 
($\sim$4fb) $t\bar tt\bar t$ cross section is needed to exclude 
(discover) a 750GeV scalar 
which decays to $t\bar t$ with 95\% ($\sim5\sigma$) confidence level (C.L.) at 14TeV LHC
with 3000fb$^{-1}$ integrated luminosity. 
The advantage of this channel is that it is $c_{NP}$-independent and 
thus can be used to measure the absolute value of
$c_t$.

To avoid these disadvantages, we notice that if the SM top-quark contribute
to the ggF production, there must be the $pp\to t\bar t\phi\to t\bar t\gamma\gamma$
process. The 
${\text{Br}}\left(\phi\to \gamma\gamma\right)$
dependence is cancelled when we take the ratio
\be
\frac{\sigma\left(pp\to t\bar t\phi\to  t\bar t\gamma\gamma\right)}
{\sigma\left(pp\to \phi\to \gamma\gamma\right)}
\propto\left|\frac{c_t}{c_t+c_{NP}}\right|^2.
\ee
This ratio only depends on $c_{NP}/c_t$, which 
tells us that if $c_{NP}=0$, the $t\bar t\gamma\gamma$ signal event
number is uniquely determined by the diphoton signal 
strength, and is a perfect 
observable to measure the size of the contribution from the 
top-quark loop. Then the relative $t\bar t\gamma\gamma$
signal strength $\mu$, which is the ratio between the
$\sigma\left(pp\to t\bar t\phi\to  t\bar t\gamma\gamma\right)$
and the cross section 
$\sigma\left(pp\to t\bar t\phi\to  t\bar t\gamma\gamma\right)_{\text{top}}$
from the rescaling of the inclusive diphoton signal strength with 
$c_{NP}=0$ assumption, is just 
\be
\left|1+\frac{c_{NP}}{c_t}\right|^{-2}.
\ee

\section{Phenomenology}
\label{sec:pheno}
\begin{figure}[t]
\includegraphics[scale=0.4,clip]{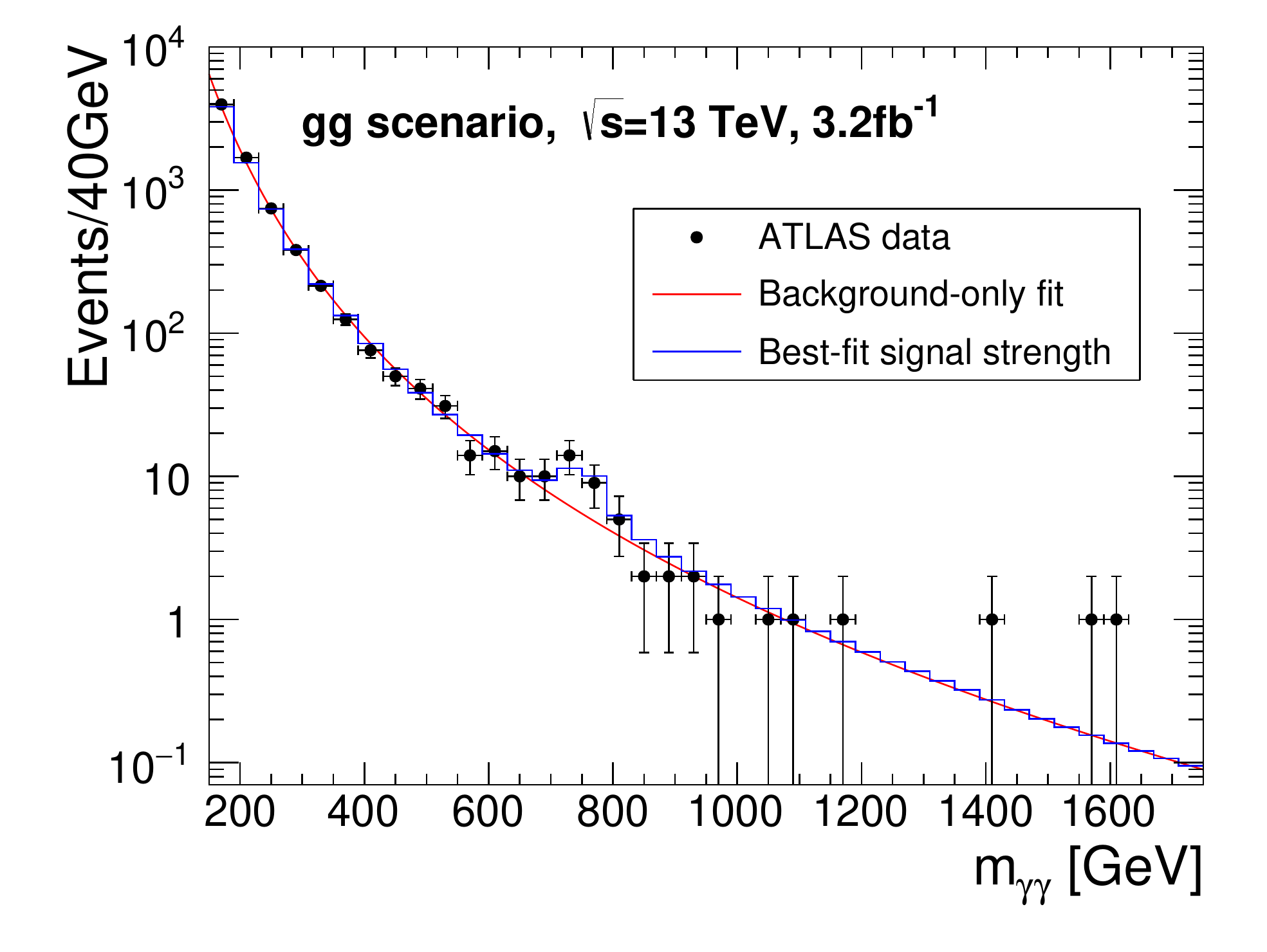}
\caption{The best-fit result of the LHC Run-II diphoton excess
with the ggF production mode \cite{Gao:2015igz}.  
\label{fig:fit} }
\end{figure}

To predict the $t\bar t\gamma\gamma$ signal strength,
we first fit the diphoton excess. In this work, we take the 
data from the ATLAS collaboration as example. We generate
parton level events using MadGraph5 \cite{Alwall:2014hca} 
with CT14llo parton distribution function (PDF) \cite{Dulat:2015mca}. 
For ggF process, $pp\to\phi+
{\text{n}} j$ events are generated to n=1. The MLM matching 
scheme is used to avoid the double counting in the parton 
showering. All parton level events are showered using 
PYTHIA6.4 with Tune Z2 parameter assignment \cite{Sjostrand:2006za,Field:2011iq}.
We use DELPHES3 to mimic the detector effects
\cite{deFavereau:2013fsa,Cacciari:2011ma}. The $b$-tagging
efficiency (and the charm and light jets mis-tagging rates) is 
tuned to be consistent with the result shown in Ref.
\cite{ATL-PHYS-PUB-2015-022}. People could find more details
of this fitting in \cite{Gao:2015igz}. We re-show the result in FIG. \ref{fig:fit}.
With this result, the unfolded signal cross section\footnote{This result
depends on the cut acceptance from the Monte Carlo (MC) simulation,
which will not be exactly. For example, if the photon identification 
rate from DELPHES is not perfectly the same to the real case, 
it will be part of the  
systematic error of the unfolding. However, most of these errors 
will be partially cancelled when we take the ratio between the 
inclusive cross sections since they also appear in the $t\bar t\phi$
MC simulation.}
\be
\sigma\left(pp\to\phi+X\right){\text{Br}}\left(\phi\to\gamma\gamma\right)
=12.3{\text{fb}}.
\ee
We calculate the 750GeV SM Higgs-like scalar ggF
cross section at 13TeV LHC to next-to-leading order (NLO) QCD level 
using MCFM7.0 \cite{Campbell:2010ff} with
CT10 PDF \cite{Lai:2010vv}. The renormalization and 
factorization scales are both set to be 375GeV. The inclusive cross section 
is 565fb. The top-pair associated production cross section 
of the SM Higgs-like scalar is calculated using MCFM7.0 
to leading order (LO) with 
MSTW2008LO PDF \cite{Martin:2009iq}, which is shown to have
a $K$-factor $\sim1$ \cite{Dittmaier:2011ti}. The renormalization and 
factorization scales are both set to be 548.2GeV. The total cross section is 
2.4fb at 13TeV LHC, 3.25fb at 14TeV LHC, and 1.00pb at 100TeV $pp$
collider. Thus we have
\bea
\sigma\left(pp\to t\bar t\phi\to t\bar t\gamma\gamma\right)&=&
\sigma\left(pp\to t\bar t\phi\right){\text{Br}}\left(\phi\to\gamma\gamma\right)\nonumber\\
&=&\frac{\sigma\left(pp\to t\bar t\phi\right)}{\sigma\left(pp\to\phi+X\right)}
\sigma\left(pp\to\phi+X\right)\nonumber\\
&&\times{\text{Br}}\left(\phi\to\gamma\gamma\right)\nonumber\\
&=&52.3\times 10^{-3}{\text{fb}}
\eea
at 13TeV LHC, 
\be
\sigma\left(pp\to t\bar t\phi\to t\bar t\gamma\gamma\right)=70.6\times 10^{-3}{\text{fb}}
\ee
at 14TeV LHC, and
\be
\sigma\left(pp\to t\bar t\phi\to t\bar t\gamma\gamma\right)=21.8{\text{fb}}
\ee
at 100TeV $pp$ collider.

In this work, we check both the dileptonic and semi-leptonic
decay modes of the top-quark pair in the $t\bar t\gamma\gamma$ events.
We add some preselection cuts on the reconstructed objects as follows:
\begin{itemize}
\item {{\bf{Photon:}} The transverse energy of the leading (subleading) 
photon should be larger than 40 (30) GeV. The pseudo-rapidity of the 
photons should satisfy
\be
|\eta^\gamma|<1.37, ~{\text{or}}~1.52<|\eta^\gamma|<2.37.
\ee
We add the isolation cut for the photon. The ratio between the summation 
of the transverse momentum of the tracks in a $\Delta R=0.4$ cone region 
around the reconstructed photon and the transverse energy of the photon 
should be smaller than 0.022 (tight selection). }
\item {{\bf{Electron:}} Electron in the pseudo-rapidity region 
\be
|\eta^e|<1.37, ~{\text{or}}~1.52<|\eta^e|<2.47
\ee
is reconstructed if its transverse momentum is larger than 25GeV. The ratio 
between the summation 
of the transverse momentum of the tracks in a $\Delta R=0.2$ cone region 
around the reconstructed electron and the transverse momentum of the electron 
should be smaller than 0.1. Electrons which are within $\Delta R<0.4$ of 
any reconstructed jet are removed from the event.}
\item {{\bf{Muon:}} Muon should satisfy 
\be
|\eta^\mu|<2.5,~ \pT^\mu>25{\text{GeV}}.
\ee
The ratio 
between the summation 
of the transverse momentum of the tracks in a $\Delta R=0.2$ cone region 
around the reconstructed electron and the transverse momentum of the electron 
should be smaller than 0.1. Muons which are within $\Delta R<0.4$ of 
any reconstructed jet are removed from the event to reduce 
the background from muons from heavy flavor decays .}
\item {{\bf{Jet:}} Jets are reconstructed using anti-$k_T$ 
algorithm with radius parameter $R=0.4$. They are accepted if
\be
|\eta^j|<2.5, ~\pT^j>25{\text{GeV}}.
\ee
In additional, $b$-jets are required to be in 
\be
|\eta^b|<2.4.
\ee}
\end{itemize}
The signal events are required to have at least one charged lepton,
two isolated hard photons and at least one $b$-tagged jet. 
Then they are separated into same-flavor 
dilepton events, $e\mu$ events and semi-leptonic events.

Some additional cuts are added for the three different signal events 
sample. The cuts are generally a combination of the SM top-pair 
cuts and high invariant mass diphoton cuts 
\cite{ATLAS-CONF-2015-081,ATLAS-CONF-2015-049,
ATLAS-CONF-2015-065}. 
First of all, the invariant mass of the leading and subleading 
photons $m_{\gamma\gamma}$ must satisfy 
\be
|m_{\gamma\gamma}-750{\text{GeV}}|<150{\text{GeV}}.
\ee
The transverse energy $E_{\text{T}}^{\gamma_1}$ 
($E_{\text{T}}^{\gamma_2}$) of the leading (subleading) photon 
must satisfy
\be
\frac{E_{\text{T}}^{\gamma_1}}{m_{\gamma\gamma}}>0.4~
\left(\frac{E_{\text{T}}^{\gamma_2}}{m_{\gamma\gamma}}>0.3\right).
\ee
\begin{itemize}
\item{{\bf{Same-flavor dilepton events:}} Events are required to have 
either exactly two opposite-sign muons or two opposite-sign electrons. 
To suppress the backgrounds from the $Z$+jets and heavy flavor decay,
the invariant mass of the dilepton system $m_{\ell\ell}$ is required to be
\be
m_{\ell\ell}>60{\text{GeV}},~|m_{\ell\ell}-m_Z|>10{\text{GeV}}.
\ee
The missing transverse energy $\met$ of the signal events must be larger 
than 30GeV.}
\item{{\bf{Semi-leptonic events:}} Events are required to have one and only
one charged lepton and at least four jets. The $\met$ and the 
transverse mass $m_{\text{T}}$ 
of the missing transverse energy and the charged lepton is required to be
\be
\met>40{\text{GeV}},~{\text{or}}~m_{\text{T}}>50{\text{GeV}}
\ee
for electron events and
\be
\met+m_{\text{T}}>60{\text{GeV}}
\ee
for muon events.}
\item{{\bf{$e\mu$ events:}} Events are required to have 
a pair of opposite-sign electron and muon. No more cut is 
added.}
\end{itemize}
All of the results of 100TeV $pp$ collider are get with the 
simple assumption that the parameters of the detector
and the cuts are the same to the LHC.

\begin{figure*}[!htb]
\includegraphics[scale=0.29,clip]{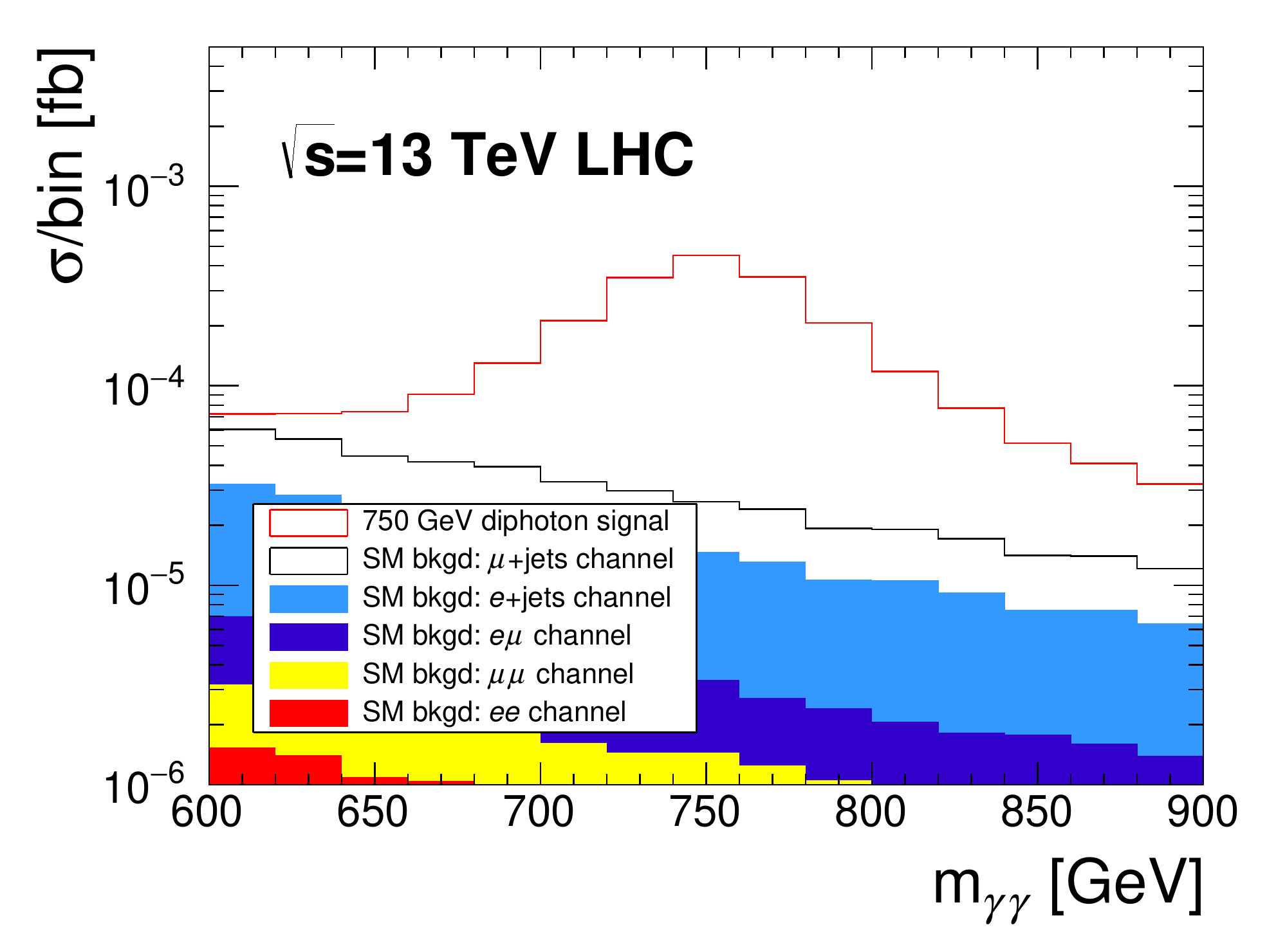}
\includegraphics[scale=0.29,clip]{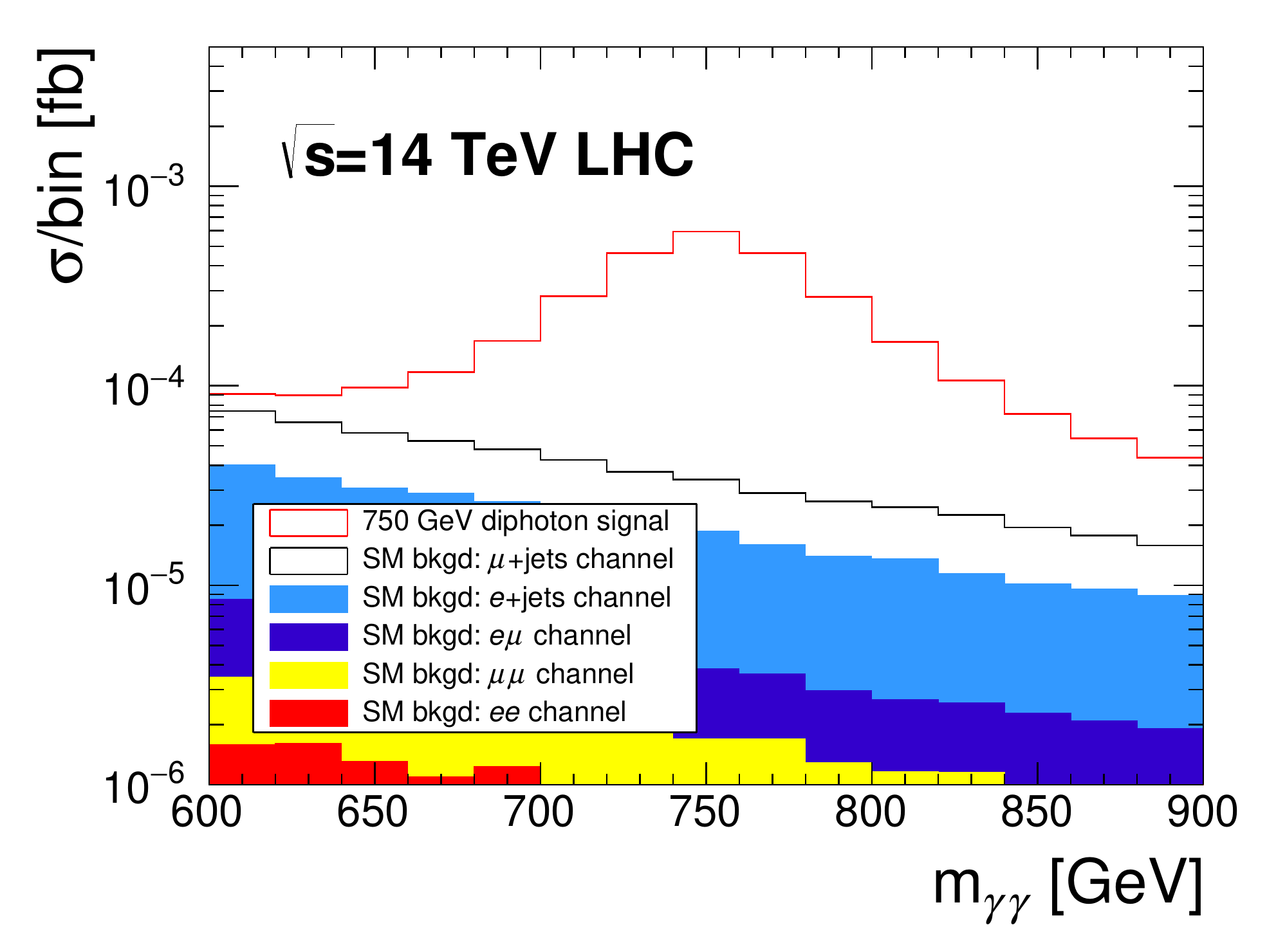}
\includegraphics[scale=0.29,clip]{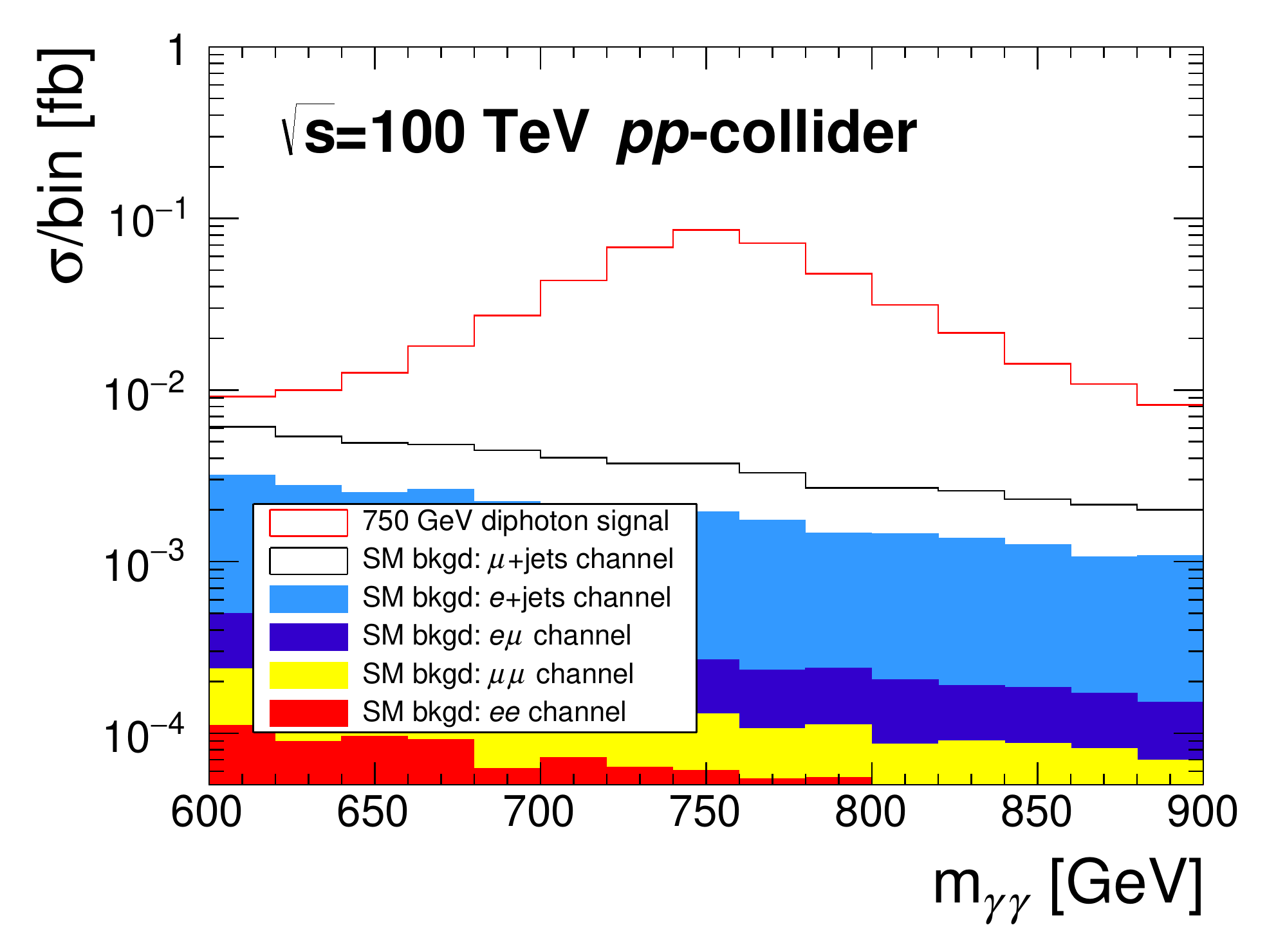}
\caption{The diphoton invariant mass distribution of both signal and 
background events after cuts. The signal strength $\mu$ is set to be
1. 
\label{fig:result} }
\end{figure*}

\begin{figure}[!htb]
\includegraphics[scale=0.4,clip]{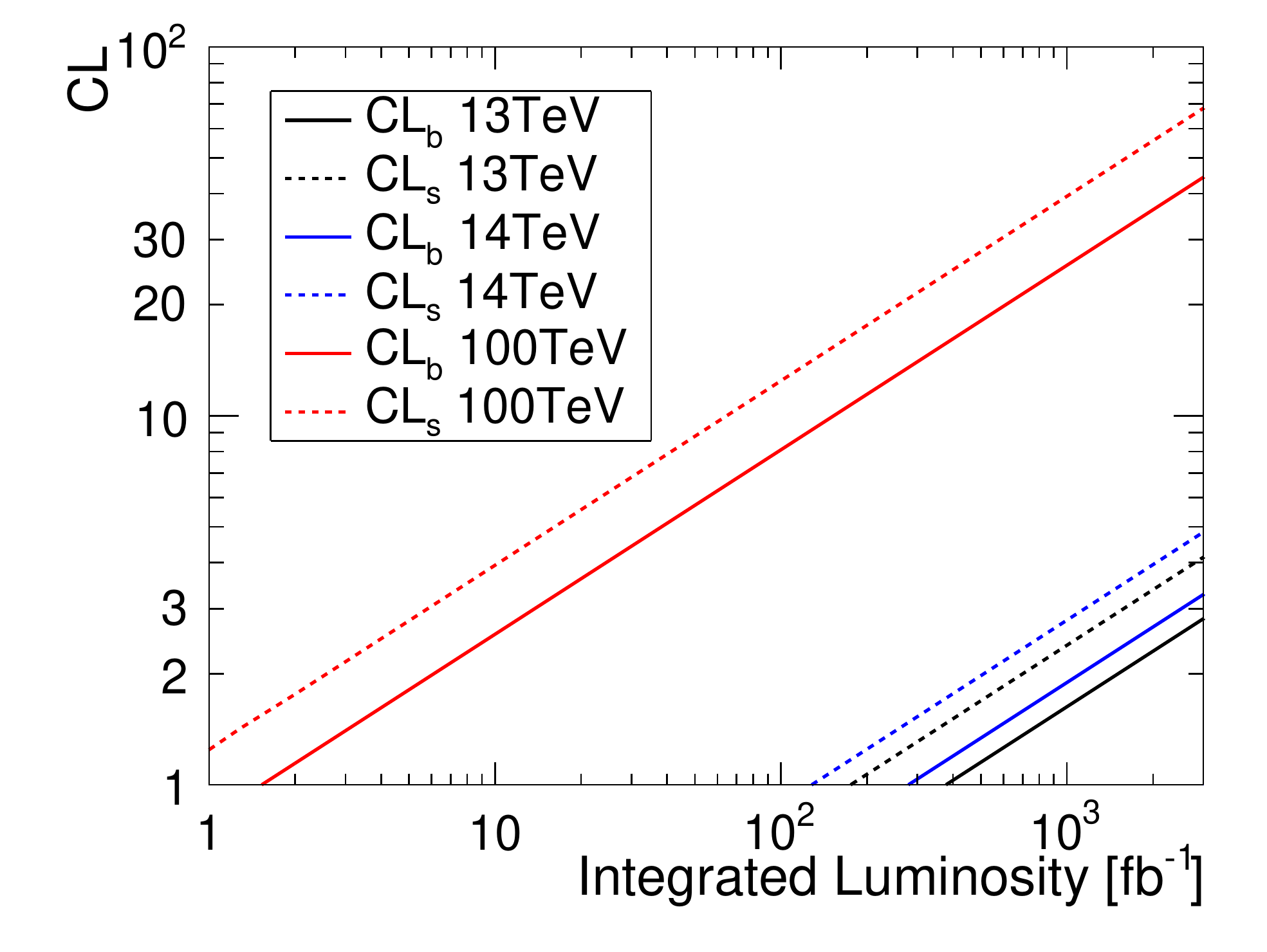}
\caption{The significant of discovery and exclusion of the top-quark-loop
only scenario at the LHC and the 100TeV $pp$ collider versus the integrated luminosity.   
\label{fig:cl} }
\end{figure}

\begin{figure}[!htb]
\includegraphics[scale=0.4,clip]{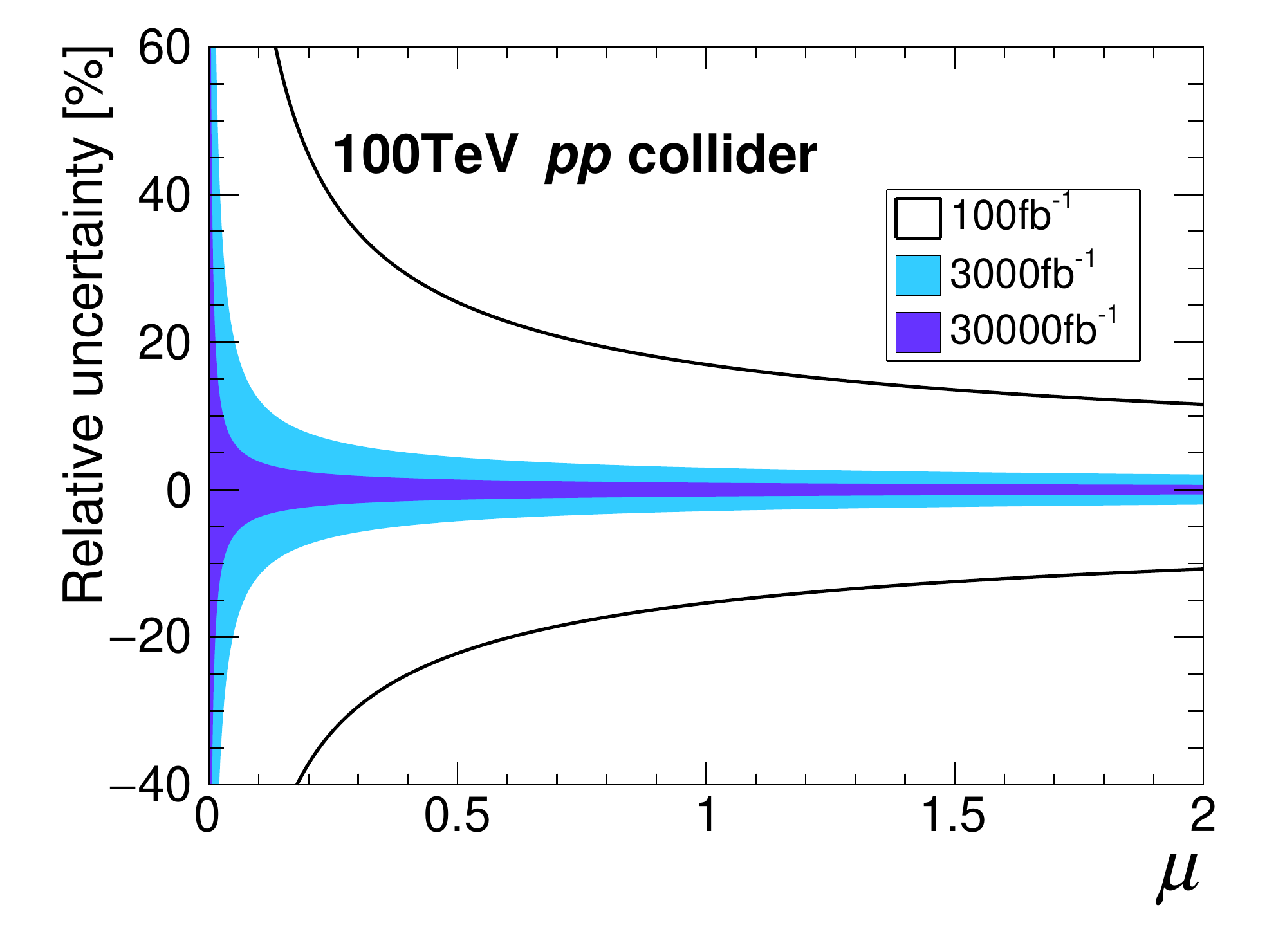}
\caption{The relative uncertainty ($\delta\mu/\mu$) of the signal 
strength at 100TeV $pp$ collider. We only consider statistical uncertainty.
\label{fig:measure} }
\end{figure}

The irreducible SM background is the $pp\to t\bar t\gamma\gamma$
process. There are some reducible SM backgrounds such as
\bea
pp&\to&t\bar t\gamma j,\nonumber\\
pp&\to&t\bar tjj,\nonumber\\
pp&\to&V+{\text{jets}}.\nonumber
\eea
However, the non-$t\bar t+X$ backgrounds would be highly suppressed 
by the cuts \cite{ATLAS-CONF-2015-049,ATLAS-CONF-2015-065}.
And the $t\bar tjj, t\bar t\gamma j$ backgrounds will be suppressed by
the mis-identification rate of a (or two) jet(s) to photon. In this preliminary
analysis, we will only consider the irreducible SM background $pp\to
t\bar t\gamma\gamma$ and neglect the irreducible backgrounds. 
Although the signal cross section is not large,
due to the extremely energetic diphoton cut, the background events number
is expected to be quite small. The results are shown in TABLE \ref{table:result}
and FIG. \ref{fig:result}.
\begin{table*}[!htb]
\caption{The signal and backgrounds cross sections after the cuts at 13 TeV LHC, 
14 TeV LHC and 100TeV $pp$ collider. The unit in this table is ab. }
\begin{center}
\begin{tabular}{c|ccccc}
\hline\hline
Channel & ~~$e$+jets channel~~ & ~~$\mu$+jets channel~~ & ~~$e\mu$ channel~~ 
& ~~$ee$ channel~~ & ~~$\mu\mu$ channel~~\\
\hline
~~13TeV background cross section (ab)~~ & $0.190$ & $0.206$ & $0.0298$ & $0.0110$ & $0.0117$ \\
13TeV signal cross section (ab) & $0.823$ & $0.873$ & $0.104$ & $0.0374$ & $0.0403$ \\
\hline
~~14TeV background cross section (ab)~~ & $0.241$ & $0.263$ & $0.0363$ & $0.0131$ & $0.0152$ \\
14TeV signal cross section (ab) & $1.11$ & $1.16$ & $0.137$ & $0.0486$ & $0.0523$ \\
\hline
~~100TeV background cross section (ab)~~ & $24.5$ & $26.0$ & $2.35$ & $0.956$ & $1.03$ \\
100TeV signal cross section (ab) & $197$ & $194$ & $19.0$ & $7.65$ & $7.06$ \\
\hline\hline
\end{tabular}
\end{center}
\label{table:result}
\end{table*}To discover the NP in the production 
process, we need to exclude
the $c_{NP}=0$ hypothesis which means $\mu=1$. 
We separate the invariant mass region 
into fifteen bins and check the exclusion significant of the signal 
with strength $\mu$ \cite{Read:2002hq,Cowan:2010js}
\be
{\text{CL}}_b\equiv\sqrt{-2\log\left[\frac{L\left(\mu\left\{s\right\}+\left\{b\right\}
|\left\{b\right\}\right)}{L\left(\left\{b\right\}|\left\{b\right\}\right)}\right]},
\ee
where $s$ and $b$ are events numbers of the signal and the background, respectively. 
If the cross sections of the signal and background are $\sigma_s$ and $\sigma_b$ 
respectively and the luminosity is $\mathcal{L}$, we have 
\be
s=\sigma_s\mathcal{L},~~b=\sigma_b\mathcal{L}.
\ee
The likelihood function is defined by 
\be
L\left(\left\{x\right\}|\left\{n\right\}\right)\equiv\prod_i\frac{x_i^{n_i}
\exp\left(-x_i\right)}{\Gamma\left(n_i+1\right)}.
\ee
People can also get the significant of confirming the top-quark loop
contribution (excluding $c_t=0$ hypothesis) which can be defined as
\be
{\text{CL}}_s\equiv\sqrt{-2\log\left[\frac{L\left(\left\{b\right\}
|\mu\left\{s\right\}+\left\{b\right\}\right)}{L\left(\mu\left\{s\right\}+\left\{b\right\}|
\mu\left\{s\right\}+\left\{b\right\}\right)}\right]}.
\ee
From FIG. \ref{fig:cl}, we find that with the full data from the 
high-luminosity (HL) LHC, a 3$\sigma$ C.L. (nearly 5$\sigma$ C.L.) exclusion
of the $c_{NP}=0$ ($c_t=0$) hypothesis
can be reached. At 100TeV $pp$ collider, the 3$\sigma$ C.L. 
exclusion of the $c_{NP}=0$ hypothesis will be 
reached with 13.8fb$^{-1}$ integrated luminosity.
To measure $\mu$ precisely, a 100TeV $pp$
collider is necessary. 
We define the uncertainty $\delta\mu$ of the signal strength by 
\be
\sqrt{-2\log\left[\frac{L\left(\left(\mu+\delta\mu\right)\left\{s\right\}+\left\{b\right\}
|\mu\left\{s\right\}+\left\{b\right\}\right)}{L\left(\mu\left\{s\right\}+\left\{b\right\}|
\mu\left\{s\right\}+\left\{b\right\}\right)}\right]}=1.
\ee
At 100TeV $pp$ collider, the signal strength $\mu$ can be measured with
about 20\% relative uncertainty with 100fb$^{-1}$ integrated luminosity,
about 3\% relative uncertainty with 3000fb$^{-1}$ integrated luminosity,
and less than 1\% relative uncertainty with 3ab$^{-1}$ integrated luminosity
(see FIG. \ref{fig:measure}). 

\section{Conclusion}
\label{sec:concl}
Recently, an excess at 750GeV in the diphoton invariant mass 
distribution is reported by ATLAS and CMS collaboration with 
the LHC Run-II data. Many works appears online to explain 
this excess. If it is confirmed by the future data, it will be the 
first particle beyond the SM discovered at high energy colliders.
And the particle physics SM must be extended. It will be very 
important to understand the production and the decay properties 
of the new particle. A ggF produced exotic scalar is one of the 
most popular explanation of this excess. In this work, we suggest 
a method to investigate the role of the SM top-quark in the 
loop-induced effective operator. 

If we are lucky, and there are new particles which contribute 
to the loop-induced effective operator, these particles might be
discovered in the searching of heavy colored particles. However,
it depends on the mass and the decay modes of the 
exotic colored particle. Comparing with directly searching for the 
exotic colored particle, our method can give a definitely answer
of the role of the SM top-quark and whether there is exotic contributions
in the production of this 750GeV resonance, with either a positive
or a null result. We show that with the HL-LHC data, 95\% 
C.L. exclusion of the top-quark-only hypothesis can be reached.
It will be a strong hint of the existence of the role of new particles in 
the production of the 750GeV resonance.

In additional, the contribution from the SM top-quark in the 
production of the 750GeV resonance can be precisely measured
with the 100TeV $pp$ collider in future \cite{CEPC-SPPCStudyGroup:2015csa,
CEPC-SPPCStudyGroup:2015esa}. With 3ab$^{-1}$ integrated
luminosity, the relative uncertainty could be smaller than 1\%.
This result, with the result from searching $pp\to t\bar t\phi\to t\bar tt\bar t$,
can help us understand the properties of the new resonance, especially 
its production. 

\section{Acknowledgments}
The work of H. Zhang is supported by the U.S. DOE under Contract No. 
DE-SC0011702.

\appendix
\bibliographystyle{apsrev}
\bibliography{draft}

\begin{thebibliography}{169}
\expandafter\ifx\csname natexlab\endcsname\relax\def\natexlab#1{#1}\fi
\expandafter\ifx\csname bibnamefont\endcsname\relax
  \def\bibnamefont#1{#1}\fi
\expandafter\ifx\csname bibfnamefont\endcsname\relax
  \def\bibfnamefont#1{#1}\fi
\expandafter\ifx\csname citenamefont\endcsname\relax
  \def\citenamefont#1{#1}\fi
\expandafter\ifx\csname url\endcsname\relax
  \def\url#1{\texttt{#1}}\fi
\expandafter\ifx\csname urlprefix\endcsname\relax\def\urlprefix{URL }\fi
\providecommand{\bibinfo}[2]{#2}
\providecommand{\eprint}[2][]{\url{#2}}

\bibitem[{ATL(2015{\natexlab{a}})}]{ATLAS-CONF-2015-081}
\bibinfo{type}{Tech. Rep.} \bibinfo{number}{ATLAS-CONF-2015-081},
  \bibinfo{institution}{CERN}, \bibinfo{address}{Geneva}
  (\bibinfo{year}{2015}{\natexlab{a}}),
  \urlprefix\url{http://cds.cern.ch/record/2114853}.

\bibitem[{CMS(2015)}]{CMS-PAS-EXO-15-004}
\bibinfo{type}{Tech. Rep.} \bibinfo{number}{CMS-PAS-EXO-15-004},
  \bibinfo{institution}{CERN}, \bibinfo{address}{Geneva}
  (\bibinfo{year}{2015}), \urlprefix\url{http://cds.cern.ch/record/2114808}.

\bibitem[{\citenamefont{Jaeckel et~al.}(2013)\citenamefont{Jaeckel, Jankowiak,
  and Spannowsky}}]{Jaeckel:2012yz}
\bibinfo{author}{\bibfnamefont{J.}~\bibnamefont{Jaeckel}},
  \bibinfo{author}{\bibfnamefont{M.}~\bibnamefont{Jankowiak}},
  \bibnamefont{and}
  \bibinfo{author}{\bibfnamefont{M.}~\bibnamefont{Spannowsky}},
  \bibinfo{journal}{Phys. Dark Univ.} \textbf{\bibinfo{volume}{2}},
  \bibinfo{pages}{111} (\bibinfo{year}{2013}), \eprint{arXiv:1212.3620}.

\bibitem[{\citenamefont{Harigaya and Nomura}(2015)}]{Harigaya:2015ezk}
\bibinfo{author}{\bibfnamefont{K.}~\bibnamefont{Harigaya}} \bibnamefont{and}
  \bibinfo{author}{\bibfnamefont{Y.}~\bibnamefont{Nomura}}
  (\bibinfo{year}{2015}), \eprint{arXiv:1512.04850}.

\bibitem[{\citenamefont{Mambrini et~al.}(2015)\citenamefont{Mambrini, Arcadi,
  and Djouadi}}]{Mambrini:2015wyu}
\bibinfo{author}{\bibfnamefont{Y.}~\bibnamefont{Mambrini}},
  \bibinfo{author}{\bibfnamefont{G.}~\bibnamefont{Arcadi}}, \bibnamefont{and}
  \bibinfo{author}{\bibfnamefont{A.}~\bibnamefont{Djouadi}}
  (\bibinfo{year}{2015}), \eprint{arXiv:1512.04913}.

\bibitem[{\citenamefont{Backovic et~al.}(2015)\citenamefont{Backovic, Mariotti,
  and Redigolo}}]{Backovic:2015fnp}
\bibinfo{author}{\bibfnamefont{M.}~\bibnamefont{Backovic}},
  \bibinfo{author}{\bibfnamefont{A.}~\bibnamefont{Mariotti}}, \bibnamefont{and}
  \bibinfo{author}{\bibfnamefont{D.}~\bibnamefont{Redigolo}}
  (\bibinfo{year}{2015}), \eprint{arXiv:1512.04917}.

\bibitem[{\citenamefont{Angelescu et~al.}(2015)\citenamefont{Angelescu,
  Djouadi, and Moreau}}]{Angelescu:2015uiz}
\bibinfo{author}{\bibfnamefont{A.}~\bibnamefont{Angelescu}},
  \bibinfo{author}{\bibfnamefont{A.}~\bibnamefont{Djouadi}}, \bibnamefont{and}
  \bibinfo{author}{\bibfnamefont{G.}~\bibnamefont{Moreau}}
  (\bibinfo{year}{2015}), \eprint{arXiv:1512.04921}.

\bibitem[{\citenamefont{Nakai et~al.}(2015)\citenamefont{Nakai, Sato, and
  Tobioka}}]{Nakai:2015ptz}
\bibinfo{author}{\bibfnamefont{Y.}~\bibnamefont{Nakai}},
  \bibinfo{author}{\bibfnamefont{R.}~\bibnamefont{Sato}}, \bibnamefont{and}
  \bibinfo{author}{\bibfnamefont{K.}~\bibnamefont{Tobioka}}
  (\bibinfo{year}{2015}), \eprint{arXiv:1512.04924}.

\bibitem[{\citenamefont{Knapen et~al.}(2015)\citenamefont{Knapen, Melia,
  Papucci, and Zurek}}]{Knapen:2015dap}
\bibinfo{author}{\bibfnamefont{S.}~\bibnamefont{Knapen}},
  \bibinfo{author}{\bibfnamefont{T.}~\bibnamefont{Melia}},
  \bibinfo{author}{\bibfnamefont{M.}~\bibnamefont{Papucci}}, \bibnamefont{and}
  \bibinfo{author}{\bibfnamefont{K.}~\bibnamefont{Zurek}}
  (\bibinfo{year}{2015}), \eprint{arXiv:1512.04928}.

\bibitem[{\citenamefont{Buttazzo et~al.}(2015)\citenamefont{Buttazzo, Greljo,
  and Marzocca}}]{Buttazzo:2015txu}
\bibinfo{author}{\bibfnamefont{D.}~\bibnamefont{Buttazzo}},
  \bibinfo{author}{\bibfnamefont{A.}~\bibnamefont{Greljo}}, \bibnamefont{and}
  \bibinfo{author}{\bibfnamefont{D.}~\bibnamefont{Marzocca}}
  (\bibinfo{year}{2015}), \eprint{arXiv:1512.04929}.

\bibitem[{\citenamefont{Pilaftsis}(2015)}]{Pilaftsis:2015ycr}
\bibinfo{author}{\bibfnamefont{A.}~\bibnamefont{Pilaftsis}}
  (\bibinfo{year}{2015}), \eprint{arXiv:1512.04931}.

\bibitem[{\citenamefont{Franceschini et~al.}(2015)\citenamefont{Franceschini,
  Giudice, Kamenik, McCullough, Pomarol, Rattazzi, Redi, Riva, Strumia, and
  Torre}}]{Franceschini:2015kwy}
\bibinfo{author}{\bibfnamefont{R.}~\bibnamefont{Franceschini}},
  \bibinfo{author}{\bibfnamefont{G.~F.} \bibnamefont{Giudice}},
  \bibinfo{author}{\bibfnamefont{J.~F.} \bibnamefont{Kamenik}},
  \bibinfo{author}{\bibfnamefont{M.}~\bibnamefont{McCullough}},
  \bibinfo{author}{\bibfnamefont{A.}~\bibnamefont{Pomarol}},
  \bibinfo{author}{\bibfnamefont{R.}~\bibnamefont{Rattazzi}},
  \bibinfo{author}{\bibfnamefont{M.}~\bibnamefont{Redi}},
  \bibinfo{author}{\bibfnamefont{F.}~\bibnamefont{Riva}},
  \bibinfo{author}{\bibfnamefont{A.}~\bibnamefont{Strumia}}, \bibnamefont{and}
  \bibinfo{author}{\bibfnamefont{R.}~\bibnamefont{Torre}}
  (\bibinfo{year}{2015}), \eprint{arXiv:1512.04933}.

\bibitem[{\citenamefont{Di~Chiara et~al.}(2015)\citenamefont{Di~Chiara,
  Marzola, and Raidal}}]{DiChiara:2015vdm}
\bibinfo{author}{\bibfnamefont{S.}~\bibnamefont{Di~Chiara}},
  \bibinfo{author}{\bibfnamefont{L.}~\bibnamefont{Marzola}}, \bibnamefont{and}
  \bibinfo{author}{\bibfnamefont{M.}~\bibnamefont{Raidal}}
  (\bibinfo{year}{2015}), \eprint{arXiv:1512.04939}.

\bibitem[{\citenamefont{McDermott et~al.}(2015)\citenamefont{McDermott, Meade,
  and Ramani}}]{McDermott:2015sck}
\bibinfo{author}{\bibfnamefont{S.~D.} \bibnamefont{McDermott}},
  \bibinfo{author}{\bibfnamefont{P.}~\bibnamefont{Meade}}, \bibnamefont{and}
  \bibinfo{author}{\bibfnamefont{H.}~\bibnamefont{Ramani}}
  (\bibinfo{year}{2015}), \eprint{arXiv:1512.05326}.

\bibitem[{\citenamefont{Ellis et~al.}(2015)\citenamefont{Ellis, Ellis,
  Quevillon, Sanz, and You}}]{Ellis:2015oso}
\bibinfo{author}{\bibfnamefont{J.}~\bibnamefont{Ellis}},
  \bibinfo{author}{\bibfnamefont{S.~A.~R.} \bibnamefont{Ellis}},
  \bibinfo{author}{\bibfnamefont{J.}~\bibnamefont{Quevillon}},
  \bibinfo{author}{\bibfnamefont{V.}~\bibnamefont{Sanz}}, \bibnamefont{and}
  \bibinfo{author}{\bibfnamefont{T.}~\bibnamefont{You}} (\bibinfo{year}{2015}),
  \eprint{arXiv:1512.05327}.

\bibitem[{\citenamefont{Low et~al.}(2015)\citenamefont{Low, Tesi, and
  Wang}}]{Low:2015qep}
\bibinfo{author}{\bibfnamefont{M.}~\bibnamefont{Low}},
  \bibinfo{author}{\bibfnamefont{A.}~\bibnamefont{Tesi}}, \bibnamefont{and}
  \bibinfo{author}{\bibfnamefont{L.-T.} \bibnamefont{Wang}}
  (\bibinfo{year}{2015}), \eprint{arXiv:1512.05328}.

\bibitem[{\citenamefont{Bellazzini et~al.}(2015)\citenamefont{Bellazzini,
  Franceschini, Sala, and Serra}}]{Bellazzini:2015nxw}
\bibinfo{author}{\bibfnamefont{B.}~\bibnamefont{Bellazzini}},
  \bibinfo{author}{\bibfnamefont{R.}~\bibnamefont{Franceschini}},
  \bibinfo{author}{\bibfnamefont{F.}~\bibnamefont{Sala}}, \bibnamefont{and}
  \bibinfo{author}{\bibfnamefont{J.}~\bibnamefont{Serra}}
  (\bibinfo{year}{2015}), \eprint{arXiv:1512.05330}.

\bibitem[{\citenamefont{Gupta et~al.}(2015)\citenamefont{Gupta, Jäger, Kats,
  Perez, and Stamou}}]{Gupta:2015zzs}
\bibinfo{author}{\bibfnamefont{R.~S.} \bibnamefont{Gupta}},
  \bibinfo{author}{\bibfnamefont{S.}~\bibnamefont{Jäger}},
  \bibinfo{author}{\bibfnamefont{Y.}~\bibnamefont{Kats}},
  \bibinfo{author}{\bibfnamefont{G.}~\bibnamefont{Perez}}, \bibnamefont{and}
  \bibinfo{author}{\bibfnamefont{E.}~\bibnamefont{Stamou}}
  (\bibinfo{year}{2015}), \eprint{arXiv:1512.05332}.

\bibitem[{\citenamefont{Petersson and Torre}(2015)}]{Petersson:2015mkr}
\bibinfo{author}{\bibfnamefont{C.}~\bibnamefont{Petersson}} \bibnamefont{and}
  \bibinfo{author}{\bibfnamefont{R.}~\bibnamefont{Torre}}
  (\bibinfo{year}{2015}), \eprint{arXiv:1512.05333}.

\bibitem[{\citenamefont{Molinaro et~al.}(2015)\citenamefont{Molinaro, Sannino,
  and Vignaroli}}]{Molinaro:2015cwg}
\bibinfo{author}{\bibfnamefont{E.}~\bibnamefont{Molinaro}},
  \bibinfo{author}{\bibfnamefont{F.}~\bibnamefont{Sannino}}, \bibnamefont{and}
  \bibinfo{author}{\bibfnamefont{N.}~\bibnamefont{Vignaroli}}
  (\bibinfo{year}{2015}), \eprint{arXiv:1512.05334}.

\bibitem[{\citenamefont{Costa et~al.}(2015)\citenamefont{Costa, Mühlleitner,
  Sampaio, and Santos}}]{Costa:2015llh}
\bibinfo{author}{\bibfnamefont{R.}~\bibnamefont{Costa}},
  \bibinfo{author}{\bibfnamefont{M.}~\bibnamefont{Mühlleitner}},
  \bibinfo{author}{\bibfnamefont{M.~O.~P.} \bibnamefont{Sampaio}},
  \bibnamefont{and} \bibinfo{author}{\bibfnamefont{R.}~\bibnamefont{Santos}}
  (\bibinfo{year}{2015}), \eprint{arXiv:1512.05355}.

\bibitem[{\citenamefont{Dutta et~al.}(2015)\citenamefont{Dutta, Gao, Ghosh,
  Gogoladze, and Li}}]{Dutta:2015wqh}
\bibinfo{author}{\bibfnamefont{B.}~\bibnamefont{Dutta}},
  \bibinfo{author}{\bibfnamefont{Y.}~\bibnamefont{Gao}},
  \bibinfo{author}{\bibfnamefont{T.}~\bibnamefont{Ghosh}},
  \bibinfo{author}{\bibfnamefont{I.}~\bibnamefont{Gogoladze}},
  \bibnamefont{and} \bibinfo{author}{\bibfnamefont{T.}~\bibnamefont{Li}}
  (\bibinfo{year}{2015}), \eprint{arXiv:1512.05439}.

\bibitem[{\citenamefont{Cao et~al.}(2015{\natexlab{a}})\citenamefont{Cao, Liu,
  Xie, Yan, and Zhang}}]{Cao:2015pto}
\bibinfo{author}{\bibfnamefont{Q.-H.} \bibnamefont{Cao}},
  \bibinfo{author}{\bibfnamefont{Y.}~\bibnamefont{Liu}},
  \bibinfo{author}{\bibfnamefont{K.-P.} \bibnamefont{Xie}},
  \bibinfo{author}{\bibfnamefont{B.}~\bibnamefont{Yan}}, \bibnamefont{and}
  \bibinfo{author}{\bibfnamefont{D.-M.} \bibnamefont{Zhang}}
  (\bibinfo{year}{2015}{\natexlab{a}}), \eprint{arXiv:1512.05542}.

\bibitem[{\citenamefont{Yamatsu}(2015)}]{Yamatsu:2015oit}
\bibinfo{author}{\bibfnamefont{N.}~\bibnamefont{Yamatsu}}
  (\bibinfo{year}{2015}), \eprint{arXiv:1512.05559}.

\bibitem[{\citenamefont{Matsuzaki and Yamawaki}(2015)}]{Matsuzaki:2015che}
\bibinfo{author}{\bibfnamefont{S.}~\bibnamefont{Matsuzaki}} \bibnamefont{and}
  \bibinfo{author}{\bibfnamefont{K.}~\bibnamefont{Yamawaki}}
  (\bibinfo{year}{2015}), \eprint{arXiv:1512.05564}.

\bibitem[{\citenamefont{Kobakhidze et~al.}(2015)\citenamefont{Kobakhidze, Wang,
  Wu, Yang, and Zhang}}]{Kobakhidze:2015ldh}
\bibinfo{author}{\bibfnamefont{A.}~\bibnamefont{Kobakhidze}},
  \bibinfo{author}{\bibfnamefont{F.}~\bibnamefont{Wang}},
  \bibinfo{author}{\bibfnamefont{L.}~\bibnamefont{Wu}},
  \bibinfo{author}{\bibfnamefont{J.~M.} \bibnamefont{Yang}}, \bibnamefont{and}
  \bibinfo{author}{\bibfnamefont{M.}~\bibnamefont{Zhang}}
  (\bibinfo{year}{2015}), \eprint{arXiv:1512.05585}.

\bibitem[{\citenamefont{Martinez et~al.}(2015)\citenamefont{Martinez, Ochoa,
  and Sierra}}]{Martinez:2015kmn}
\bibinfo{author}{\bibfnamefont{R.}~\bibnamefont{Martinez}},
  \bibinfo{author}{\bibfnamefont{F.}~\bibnamefont{Ochoa}}, \bibnamefont{and}
  \bibinfo{author}{\bibfnamefont{C.~F.} \bibnamefont{Sierra}}
  (\bibinfo{year}{2015}), \eprint{arXiv:1512.05617}.

\bibitem[{\citenamefont{Cox et~al.}(2015)\citenamefont{Cox, Medina, Ray, and
  Spray}}]{Cox:2015ckc}
\bibinfo{author}{\bibfnamefont{P.}~\bibnamefont{Cox}},
  \bibinfo{author}{\bibfnamefont{A.~D.} \bibnamefont{Medina}},
  \bibinfo{author}{\bibfnamefont{T.~S.} \bibnamefont{Ray}}, \bibnamefont{and}
  \bibinfo{author}{\bibfnamefont{A.}~\bibnamefont{Spray}}
  (\bibinfo{year}{2015}), \eprint{arXiv:1512.05618}.

\bibitem[{\citenamefont{Becirevic et~al.}(2015)\citenamefont{Becirevic,
  Bertuzzo, Sumensari, and Funchal}}]{Becirevic:2015fmu}
\bibinfo{author}{\bibfnamefont{D.}~\bibnamefont{Becirevic}},
  \bibinfo{author}{\bibfnamefont{E.}~\bibnamefont{Bertuzzo}},
  \bibinfo{author}{\bibfnamefont{O.}~\bibnamefont{Sumensari}},
  \bibnamefont{and} \bibinfo{author}{\bibfnamefont{R.~Z.}
  \bibnamefont{Funchal}} (\bibinfo{year}{2015}), \eprint{arXiv:1512.05623}.

\bibitem[{\citenamefont{No et~al.}(2015)\citenamefont{No, Sanz, and
  Setford}}]{No:2015bsn}
\bibinfo{author}{\bibfnamefont{J.~M.} \bibnamefont{No}},
  \bibinfo{author}{\bibfnamefont{V.}~\bibnamefont{Sanz}}, \bibnamefont{and}
  \bibinfo{author}{\bibfnamefont{J.}~\bibnamefont{Setford}}
  (\bibinfo{year}{2015}), \eprint{arXiv:1512.05700}.

\bibitem[{\citenamefont{Demidov and Gorbunov}(2015)}]{Demidov:2015zqn}
\bibinfo{author}{\bibfnamefont{S.~V.} \bibnamefont{Demidov}} \bibnamefont{and}
  \bibinfo{author}{\bibfnamefont{D.~S.} \bibnamefont{Gorbunov}}
  (\bibinfo{year}{2015}), \eprint{arXiv:1512.05723}.

\bibitem[{\citenamefont{Gopalakrishna et~al.}(2015)\citenamefont{Gopalakrishna,
  Mukherjee, and Sadhukhan}}]{Gopalakrishna:2015dkt}
\bibinfo{author}{\bibfnamefont{S.}~\bibnamefont{Gopalakrishna}},
  \bibinfo{author}{\bibfnamefont{T.~S.} \bibnamefont{Mukherjee}},
  \bibnamefont{and} \bibinfo{author}{\bibfnamefont{S.}~\bibnamefont{Sadhukhan}}
  (\bibinfo{year}{2015}), \eprint{arXiv:1512.05731}.

\bibitem[{\citenamefont{Chao et~al.}(2015)\citenamefont{Chao, Huo, and
  Yu}}]{Chao:2015ttq}
\bibinfo{author}{\bibfnamefont{W.}~\bibnamefont{Chao}},
  \bibinfo{author}{\bibfnamefont{R.}~\bibnamefont{Huo}}, \bibnamefont{and}
  \bibinfo{author}{\bibfnamefont{J.-H.} \bibnamefont{Yu}}
  (\bibinfo{year}{2015}), \eprint{arXiv:1512.05738}.

\bibitem[{\citenamefont{Fichet et~al.}(2015)\citenamefont{Fichet, von
  Gersdorff, and Royon}}]{Fichet:2015vvy}
\bibinfo{author}{\bibfnamefont{S.}~\bibnamefont{Fichet}},
  \bibinfo{author}{\bibfnamefont{G.}~\bibnamefont{von Gersdorff}},
  \bibnamefont{and} \bibinfo{author}{\bibfnamefont{C.}~\bibnamefont{Royon}}
  (\bibinfo{year}{2015}), \eprint{arXiv:1512.05751}.

\bibitem[{\citenamefont{Bian et~al.}(2015)\citenamefont{Bian, Chen, Liu, and
  Shu}}]{Bian:2015kjt}
\bibinfo{author}{\bibfnamefont{L.}~\bibnamefont{Bian}},
  \bibinfo{author}{\bibfnamefont{N.}~\bibnamefont{Chen}},
  \bibinfo{author}{\bibfnamefont{D.}~\bibnamefont{Liu}}, \bibnamefont{and}
  \bibinfo{author}{\bibfnamefont{J.}~\bibnamefont{Shu}} (\bibinfo{year}{2015}),
  \eprint{arXiv:1512.05759}.

\bibitem[{\citenamefont{Chakrabortty et~al.}(2015)\citenamefont{Chakrabortty,
  Choudhury, Ghosh, Mondal, and Srivastava}}]{Chakrabortty:2015hff}
\bibinfo{author}{\bibfnamefont{J.}~\bibnamefont{Chakrabortty}},
  \bibinfo{author}{\bibfnamefont{A.}~\bibnamefont{Choudhury}},
  \bibinfo{author}{\bibfnamefont{P.}~\bibnamefont{Ghosh}},
  \bibinfo{author}{\bibfnamefont{S.}~\bibnamefont{Mondal}}, \bibnamefont{and}
  \bibinfo{author}{\bibfnamefont{T.}~\bibnamefont{Srivastava}}
  (\bibinfo{year}{2015}), \eprint{arXiv:1512.05767}.

\bibitem[{\citenamefont{Ahmed et~al.}(2015)\citenamefont{Ahmed, Dillon,
  Grzadkowski, Gunion, and Jiang}}]{Ahmed:2015uqt}
\bibinfo{author}{\bibfnamefont{A.}~\bibnamefont{Ahmed}},
  \bibinfo{author}{\bibfnamefont{B.~M.} \bibnamefont{Dillon}},
  \bibinfo{author}{\bibfnamefont{B.}~\bibnamefont{Grzadkowski}},
  \bibinfo{author}{\bibfnamefont{J.~F.} \bibnamefont{Gunion}},
  \bibnamefont{and} \bibinfo{author}{\bibfnamefont{Y.}~\bibnamefont{Jiang}}
  (\bibinfo{year}{2015}), \eprint{arXiv:1512.05771}.

\bibitem[{\citenamefont{Agrawal et~al.}(2015)\citenamefont{Agrawal, Fan,
  Heidenreich, Reece, and Strassler}}]{Agrawal:2015dbf}
\bibinfo{author}{\bibfnamefont{P.}~\bibnamefont{Agrawal}},
  \bibinfo{author}{\bibfnamefont{J.}~\bibnamefont{Fan}},
  \bibinfo{author}{\bibfnamefont{B.}~\bibnamefont{Heidenreich}},
  \bibinfo{author}{\bibfnamefont{M.}~\bibnamefont{Reece}}, \bibnamefont{and}
  \bibinfo{author}{\bibfnamefont{M.}~\bibnamefont{Strassler}}
  (\bibinfo{year}{2015}), \eprint{arXiv:1512.05775}.

\bibitem[{\citenamefont{Csaki et~al.}(2015)\citenamefont{Csaki, Hubisz, and
  Terning}}]{Csaki:2015vek}
\bibinfo{author}{\bibfnamefont{C.}~\bibnamefont{Csaki}},
  \bibinfo{author}{\bibfnamefont{J.}~\bibnamefont{Hubisz}}, \bibnamefont{and}
  \bibinfo{author}{\bibfnamefont{J.}~\bibnamefont{Terning}}
  (\bibinfo{year}{2015}), \eprint{arXiv:1512.05776}.

\bibitem[{\citenamefont{Falkowski et~al.}(2015)\citenamefont{Falkowski, Slone,
  and Volansky}}]{Falkowski:2015swt}
\bibinfo{author}{\bibfnamefont{A.}~\bibnamefont{Falkowski}},
  \bibinfo{author}{\bibfnamefont{O.}~\bibnamefont{Slone}}, \bibnamefont{and}
  \bibinfo{author}{\bibfnamefont{T.}~\bibnamefont{Volansky}}
  (\bibinfo{year}{2015}), \eprint{arXiv:1512.05777}.

\bibitem[{\citenamefont{Aloni et~al.}(2015)\citenamefont{Aloni, Blum, Dery,
  Efrati, and Nir}}]{Aloni:2015mxa}
\bibinfo{author}{\bibfnamefont{D.}~\bibnamefont{Aloni}},
  \bibinfo{author}{\bibfnamefont{K.}~\bibnamefont{Blum}},
  \bibinfo{author}{\bibfnamefont{A.}~\bibnamefont{Dery}},
  \bibinfo{author}{\bibfnamefont{A.}~\bibnamefont{Efrati}}, \bibnamefont{and}
  \bibinfo{author}{\bibfnamefont{Y.}~\bibnamefont{Nir}} (\bibinfo{year}{2015}),
  \eprint{arXiv:1512.05778}.

\bibitem[{\citenamefont{Bai et~al.}(2015)\citenamefont{Bai, Berger, and
  Lu}}]{Bai:2015nbs}
\bibinfo{author}{\bibfnamefont{Y.}~\bibnamefont{Bai}},
  \bibinfo{author}{\bibfnamefont{J.}~\bibnamefont{Berger}}, \bibnamefont{and}
  \bibinfo{author}{\bibfnamefont{R.}~\bibnamefont{Lu}} (\bibinfo{year}{2015}),
  \eprint{arXiv:1512.05779}.

\bibitem[{\citenamefont{Gabrielli et~al.}(2015)\citenamefont{Gabrielli,
  Kannike, Mele, Raidal, Spethmann, and Veermäe}}]{Gabrielli:2015dhk}
\bibinfo{author}{\bibfnamefont{E.}~\bibnamefont{Gabrielli}},
  \bibinfo{author}{\bibfnamefont{K.}~\bibnamefont{Kannike}},
  \bibinfo{author}{\bibfnamefont{B.}~\bibnamefont{Mele}},
  \bibinfo{author}{\bibfnamefont{M.}~\bibnamefont{Raidal}},
  \bibinfo{author}{\bibfnamefont{C.}~\bibnamefont{Spethmann}},
  \bibnamefont{and} \bibinfo{author}{\bibfnamefont{H.}~\bibnamefont{Veermäe}}
  (\bibinfo{year}{2015}), \eprint{arXiv:1512.05961}.

\bibitem[{\citenamefont{Benbrik et~al.}(2015)\citenamefont{Benbrik, Chen, and
  Nomura}}]{Benbrik:2015fyz}
\bibinfo{author}{\bibfnamefont{R.}~\bibnamefont{Benbrik}},
  \bibinfo{author}{\bibfnamefont{C.-H.} \bibnamefont{Chen}}, \bibnamefont{and}
  \bibinfo{author}{\bibfnamefont{T.}~\bibnamefont{Nomura}}
  (\bibinfo{year}{2015}), \eprint{arXiv:1512.06028}.

\bibitem[{\citenamefont{Kim et~al.}(2015{\natexlab{a}})\citenamefont{Kim,
  Reuter, Rolbiecki, and de~Austri}}]{Kim:2015ron}
\bibinfo{author}{\bibfnamefont{J.~S.} \bibnamefont{Kim}},
  \bibinfo{author}{\bibfnamefont{J.}~\bibnamefont{Reuter}},
  \bibinfo{author}{\bibfnamefont{K.}~\bibnamefont{Rolbiecki}},
  \bibnamefont{and} \bibinfo{author}{\bibfnamefont{R.~R.}
  \bibnamefont{de~Austri}} (\bibinfo{year}{2015}{\natexlab{a}}),
  \eprint{arXiv:1512.06083}.

\bibitem[{\citenamefont{Alves et~al.}(2015)\citenamefont{Alves, Dias, and
  Sinha}}]{Alves:2015jgx}
\bibinfo{author}{\bibfnamefont{A.}~\bibnamefont{Alves}},
  \bibinfo{author}{\bibfnamefont{A.~G.} \bibnamefont{Dias}}, \bibnamefont{and}
  \bibinfo{author}{\bibfnamefont{K.}~\bibnamefont{Sinha}}
  (\bibinfo{year}{2015}), \eprint{arXiv:1512.06091}.

\bibitem[{\citenamefont{Megias et~al.}(2015)\citenamefont{Megias, Pujolas, and
  Quiros}}]{Megias:2015ory}
\bibinfo{author}{\bibfnamefont{E.}~\bibnamefont{Megias}},
  \bibinfo{author}{\bibfnamefont{O.}~\bibnamefont{Pujolas}}, \bibnamefont{and}
  \bibinfo{author}{\bibfnamefont{M.}~\bibnamefont{Quiros}}
  (\bibinfo{year}{2015}), \eprint{arXiv:1512.06106}.

\bibitem[{\citenamefont{Carpenter et~al.}(2015)\citenamefont{Carpenter,
  Colburn, and Goodman}}]{Carpenter:2015ucu}
\bibinfo{author}{\bibfnamefont{L.~M.} \bibnamefont{Carpenter}},
  \bibinfo{author}{\bibfnamefont{R.}~\bibnamefont{Colburn}}, \bibnamefont{and}
  \bibinfo{author}{\bibfnamefont{J.}~\bibnamefont{Goodman}}
  (\bibinfo{year}{2015}), \eprint{arXiv:1512.06107}.

\bibitem[{\citenamefont{Chao}(2015{\natexlab{a}})}]{Chao:2015nsm}
\bibinfo{author}{\bibfnamefont{W.}~\bibnamefont{Chao}}
  (\bibinfo{year}{2015}{\natexlab{a}}), \eprint{arXiv:1512.06297}.

\bibitem[{\citenamefont{Arun and Saha}(2015)}]{Arun:2015ubr}
\bibinfo{author}{\bibfnamefont{M.~T.} \bibnamefont{Arun}} \bibnamefont{and}
  \bibinfo{author}{\bibfnamefont{P.}~\bibnamefont{Saha}}
  (\bibinfo{year}{2015}), \eprint{arXiv:1512.06335}.

\bibitem[{\citenamefont{Han et~al.}(2015{\natexlab{a}})\citenamefont{Han, Lee,
  Park, and Sanz}}]{Han:2015cty}
\bibinfo{author}{\bibfnamefont{C.}~\bibnamefont{Han}},
  \bibinfo{author}{\bibfnamefont{H.~M.} \bibnamefont{Lee}},
  \bibinfo{author}{\bibfnamefont{M.}~\bibnamefont{Park}}, \bibnamefont{and}
  \bibinfo{author}{\bibfnamefont{V.}~\bibnamefont{Sanz}}
  (\bibinfo{year}{2015}{\natexlab{a}}), \eprint{arXiv:1512.06376}.

\bibitem[{\citenamefont{Chang}(2015)}]{Chang:2015bzc}
\bibinfo{author}{\bibfnamefont{S.}~\bibnamefont{Chang}} (\bibinfo{year}{2015}),
  \eprint{arXiv:1512.06426}.

\bibitem[{\citenamefont{Chakraborty and Kundu}(2015)}]{Chakraborty:2015jvs}
\bibinfo{author}{\bibfnamefont{I.}~\bibnamefont{Chakraborty}} \bibnamefont{and}
  \bibinfo{author}{\bibfnamefont{A.}~\bibnamefont{Kundu}}
  (\bibinfo{year}{2015}), \eprint{arXiv:1512.06508}.

\bibitem[{\citenamefont{Ding et~al.}(2015)\citenamefont{Ding, Huang, Li, and
  Zhu}}]{Ding:2015rxx}
\bibinfo{author}{\bibfnamefont{R.}~\bibnamefont{Ding}},
  \bibinfo{author}{\bibfnamefont{L.}~\bibnamefont{Huang}},
  \bibinfo{author}{\bibfnamefont{T.}~\bibnamefont{Li}}, \bibnamefont{and}
  \bibinfo{author}{\bibfnamefont{B.}~\bibnamefont{Zhu}} (\bibinfo{year}{2015}),
  \eprint{arXiv:1512.06560}.

\bibitem[{\citenamefont{Han et~al.}(2015{\natexlab{b}})\citenamefont{Han, Wang,
  and Zheng}}]{Han:2015dlp}
\bibinfo{author}{\bibfnamefont{H.}~\bibnamefont{Han}},
  \bibinfo{author}{\bibfnamefont{S.}~\bibnamefont{Wang}}, \bibnamefont{and}
  \bibinfo{author}{\bibfnamefont{S.}~\bibnamefont{Zheng}}
  (\bibinfo{year}{2015}{\natexlab{b}}), \eprint{arXiv:1512.06562}.

\bibitem[{\citenamefont{Han and Wang}(2015)}]{Han:2015qqj}
\bibinfo{author}{\bibfnamefont{X.-F.} \bibnamefont{Han}} \bibnamefont{and}
  \bibinfo{author}{\bibfnamefont{L.}~\bibnamefont{Wang}}
  (\bibinfo{year}{2015}), \eprint{arXiv:1512.06587}.

\bibitem[{\citenamefont{Luo et~al.}(2015)\citenamefont{Luo, Wang, Xu, Zhang,
  and Zhu}}]{Luo:2015yio}
\bibinfo{author}{\bibfnamefont{M.-x.} \bibnamefont{Luo}},
  \bibinfo{author}{\bibfnamefont{K.}~\bibnamefont{Wang}},
  \bibinfo{author}{\bibfnamefont{T.}~\bibnamefont{Xu}},
  \bibinfo{author}{\bibfnamefont{L.}~\bibnamefont{Zhang}}, \bibnamefont{and}
  \bibinfo{author}{\bibfnamefont{G.}~\bibnamefont{Zhu}} (\bibinfo{year}{2015}),
  \eprint{arXiv:1512.06670}.

\bibitem[{\citenamefont{Chang et~al.}(2015)\citenamefont{Chang, Cheung, and
  Lu}}]{Chang:2015sdy}
\bibinfo{author}{\bibfnamefont{J.}~\bibnamefont{Chang}},
  \bibinfo{author}{\bibfnamefont{K.}~\bibnamefont{Cheung}}, \bibnamefont{and}
  \bibinfo{author}{\bibfnamefont{C.-T.} \bibnamefont{Lu}}
  (\bibinfo{year}{2015}), \eprint{arXiv:1512.06671}.

\bibitem[{\citenamefont{Bardhan et~al.}(2015)\citenamefont{Bardhan, Bhatia,
  Chakraborty, Maitra, Raychaudhuri, and Samui}}]{Bardhan:2015hcr}
\bibinfo{author}{\bibfnamefont{D.}~\bibnamefont{Bardhan}},
  \bibinfo{author}{\bibfnamefont{D.}~\bibnamefont{Bhatia}},
  \bibinfo{author}{\bibfnamefont{A.}~\bibnamefont{Chakraborty}},
  \bibinfo{author}{\bibfnamefont{U.}~\bibnamefont{Maitra}},
  \bibinfo{author}{\bibfnamefont{S.}~\bibnamefont{Raychaudhuri}},
  \bibnamefont{and} \bibinfo{author}{\bibfnamefont{T.}~\bibnamefont{Samui}}
  (\bibinfo{year}{2015}), \eprint{arXiv:1512.06674}.

\bibitem[{\citenamefont{Feng et~al.}(2015)\citenamefont{Feng, Li, Zhang, and
  Zhao}}]{Feng:2015wil}
\bibinfo{author}{\bibfnamefont{T.-F.} \bibnamefont{Feng}},
  \bibinfo{author}{\bibfnamefont{X.-Q.} \bibnamefont{Li}},
  \bibinfo{author}{\bibfnamefont{H.-B.} \bibnamefont{Zhang}}, \bibnamefont{and}
  \bibinfo{author}{\bibfnamefont{S.-M.} \bibnamefont{Zhao}}
  (\bibinfo{year}{2015}), \eprint{arXiv:1512.06696}.

\bibitem[{\citenamefont{Wang et~al.}(2015{\natexlab{a}})\citenamefont{Wang, Wu,
  Yang, and Zhang}}]{Wang:2015kuj}
\bibinfo{author}{\bibfnamefont{F.}~\bibnamefont{Wang}},
  \bibinfo{author}{\bibfnamefont{L.}~\bibnamefont{Wu}},
  \bibinfo{author}{\bibfnamefont{J.~M.} \bibnamefont{Yang}}, \bibnamefont{and}
  \bibinfo{author}{\bibfnamefont{M.}~\bibnamefont{Zhang}}
  (\bibinfo{year}{2015}{\natexlab{a}}), \eprint{arXiv:1512.06715}.

\bibitem[{\citenamefont{Antipin et~al.}(2015)\citenamefont{Antipin, Mojaza, and
  Sannino}}]{Antipin:2015kgh}
\bibinfo{author}{\bibfnamefont{O.}~\bibnamefont{Antipin}},
  \bibinfo{author}{\bibfnamefont{M.}~\bibnamefont{Mojaza}}, \bibnamefont{and}
  \bibinfo{author}{\bibfnamefont{F.}~\bibnamefont{Sannino}}
  (\bibinfo{year}{2015}), \eprint{arXiv:1512.06708}.

\bibitem[{\citenamefont{Cao et~al.}(2015{\natexlab{b}})\citenamefont{Cao, Han,
  Shang, Su, Yang, and Zhang}}]{Cao:2015twy}
\bibinfo{author}{\bibfnamefont{J.}~\bibnamefont{Cao}},
  \bibinfo{author}{\bibfnamefont{C.}~\bibnamefont{Han}},
  \bibinfo{author}{\bibfnamefont{L.}~\bibnamefont{Shang}},
  \bibinfo{author}{\bibfnamefont{W.}~\bibnamefont{Su}},
  \bibinfo{author}{\bibfnamefont{J.~M.} \bibnamefont{Yang}}, \bibnamefont{and}
  \bibinfo{author}{\bibfnamefont{Y.}~\bibnamefont{Zhang}}
  (\bibinfo{year}{2015}{\natexlab{b}}), \eprint{arXiv:1512.06728}.

\bibitem[{\citenamefont{Huang et~al.}(2015{\natexlab{a}})\citenamefont{Huang,
  Li, Liu, and Wang}}]{Huang:2015evq}
\bibinfo{author}{\bibfnamefont{F.~P.} \bibnamefont{Huang}},
  \bibinfo{author}{\bibfnamefont{C.~S.} \bibnamefont{Li}},
  \bibinfo{author}{\bibfnamefont{Z.~L.} \bibnamefont{Liu}}, \bibnamefont{and}
  \bibinfo{author}{\bibfnamefont{Y.}~\bibnamefont{Wang}}
  (\bibinfo{year}{2015}{\natexlab{a}}), \eprint{arXiv:1512.06732}.

\bibitem[{\citenamefont{Liao and Zheng}(2015)}]{Liao:2015tow}
\bibinfo{author}{\bibfnamefont{W.}~\bibnamefont{Liao}} \bibnamefont{and}
  \bibinfo{author}{\bibfnamefont{H.-q.} \bibnamefont{Zheng}}
  (\bibinfo{year}{2015}), \eprint{arXiv:1512.06741}.

\bibitem[{\citenamefont{Heckman}(2015)}]{Heckman:2015kqk}
\bibinfo{author}{\bibfnamefont{J.~J.} \bibnamefont{Heckman}}
  (\bibinfo{year}{2015}), \eprint{arXiv:1512.06773}.

\bibitem[{\citenamefont{Dhuria and Goswami}(2015)}]{Dhuria:2015ufo}
\bibinfo{author}{\bibfnamefont{M.}~\bibnamefont{Dhuria}} \bibnamefont{and}
  \bibinfo{author}{\bibfnamefont{G.}~\bibnamefont{Goswami}}
  (\bibinfo{year}{2015}), \eprint{arXiv:1512.06782}.

\bibitem[{\citenamefont{Bi et~al.}(2015{\natexlab{a}})\citenamefont{Bi, Xiang,
  Yin, and Yu}}]{Bi:2015uqd}
\bibinfo{author}{\bibfnamefont{X.-J.} \bibnamefont{Bi}},
  \bibinfo{author}{\bibfnamefont{Q.-F.} \bibnamefont{Xiang}},
  \bibinfo{author}{\bibfnamefont{P.-F.} \bibnamefont{Yin}}, \bibnamefont{and}
  \bibinfo{author}{\bibfnamefont{Z.-H.} \bibnamefont{Yu}}
  (\bibinfo{year}{2015}{\natexlab{a}}), \eprint{arXiv:1512.06787}.

\bibitem[{\citenamefont{Kim et~al.}(2015{\natexlab{b}})\citenamefont{Kim,
  Rolbiecki, and de~Austri}}]{Kim:2015ksf}
\bibinfo{author}{\bibfnamefont{J.~S.} \bibnamefont{Kim}},
  \bibinfo{author}{\bibfnamefont{K.}~\bibnamefont{Rolbiecki}},
  \bibnamefont{and} \bibinfo{author}{\bibfnamefont{R.~R.}
  \bibnamefont{de~Austri}} (\bibinfo{year}{2015}{\natexlab{b}}),
  \eprint{arXiv:1512.06797}.

\bibitem[{\citenamefont{Berthier et~al.}(2015)\citenamefont{Berthier, Cline,
  Shepherd, and Trott}}]{Berthier:2015vbb}
\bibinfo{author}{\bibfnamefont{L.}~\bibnamefont{Berthier}},
  \bibinfo{author}{\bibfnamefont{J.~M.} \bibnamefont{Cline}},
  \bibinfo{author}{\bibfnamefont{W.}~\bibnamefont{Shepherd}}, \bibnamefont{and}
  \bibinfo{author}{\bibfnamefont{M.}~\bibnamefont{Trott}}
  (\bibinfo{year}{2015}), \eprint{arXiv:1512.06799}.

\bibitem[{\citenamefont{Cho et~al.}(2015)\citenamefont{Cho, Kim, Kong, Lim,
  Matchev, Park, and Park}}]{Cho:2015nxy}
\bibinfo{author}{\bibfnamefont{W.~S.} \bibnamefont{Cho}},
  \bibinfo{author}{\bibfnamefont{D.}~\bibnamefont{Kim}},
  \bibinfo{author}{\bibfnamefont{K.}~\bibnamefont{Kong}},
  \bibinfo{author}{\bibfnamefont{S.~H.} \bibnamefont{Lim}},
  \bibinfo{author}{\bibfnamefont{K.~T.} \bibnamefont{Matchev}},
  \bibinfo{author}{\bibfnamefont{J.-C.} \bibnamefont{Park}}, \bibnamefont{and}
  \bibinfo{author}{\bibfnamefont{M.}~\bibnamefont{Park}}
  (\bibinfo{year}{2015}), \eprint{arXiv:1512.06824}.

\bibitem[{\citenamefont{Cline and Liu}(2015)}]{Cline:2015msi}
\bibinfo{author}{\bibfnamefont{J.~M.} \bibnamefont{Cline}} \bibnamefont{and}
  \bibinfo{author}{\bibfnamefont{Z.}~\bibnamefont{Liu}} (\bibinfo{year}{2015}),
  \eprint{arXiv:1512.06827}.

\bibitem[{\citenamefont{Chala et~al.}(2015)\citenamefont{Chala, Duerr,
  Kahlhoefer, and Schmidt-Hoberg}}]{Chala:2015cev}
\bibinfo{author}{\bibfnamefont{M.}~\bibnamefont{Chala}},
  \bibinfo{author}{\bibfnamefont{M.}~\bibnamefont{Duerr}},
  \bibinfo{author}{\bibfnamefont{F.}~\bibnamefont{Kahlhoefer}},
  \bibnamefont{and}
  \bibinfo{author}{\bibfnamefont{K.}~\bibnamefont{Schmidt-Hoberg}}
  (\bibinfo{year}{2015}), \eprint{arXiv:1512.06833}.

\bibitem[{\citenamefont{Barducci et~al.}(2015)\citenamefont{Barducci, Goudelis,
  Kulkarni, and Sengupta}}]{Barducci:2015gtd}
\bibinfo{author}{\bibfnamefont{D.}~\bibnamefont{Barducci}},
  \bibinfo{author}{\bibfnamefont{A.}~\bibnamefont{Goudelis}},
  \bibinfo{author}{\bibfnamefont{S.}~\bibnamefont{Kulkarni}}, \bibnamefont{and}
  \bibinfo{author}{\bibfnamefont{D.}~\bibnamefont{Sengupta}}
  (\bibinfo{year}{2015}), \eprint{arXiv:1512.06842}.

\bibitem[{\citenamefont{Boucenna et~al.}(2015)\citenamefont{Boucenna, Morisi,
  and Vicente}}]{Boucenna:2015pav}
\bibinfo{author}{\bibfnamefont{S.~M.} \bibnamefont{Boucenna}},
  \bibinfo{author}{\bibfnamefont{S.}~\bibnamefont{Morisi}}, \bibnamefont{and}
  \bibinfo{author}{\bibfnamefont{A.}~\bibnamefont{Vicente}}
  (\bibinfo{year}{2015}), \eprint{arXiv:1512.06878}.

\bibitem[{\citenamefont{Murphy}(2015)}]{Murphy:2015kag}
\bibinfo{author}{\bibfnamefont{C.~W.} \bibnamefont{Murphy}}
  (\bibinfo{year}{2015}), \eprint{arXiv:1512.06976}.

\bibitem[{\citenamefont{Hernández and Nisandzic}(2015)}]{Hernandez:2015ywg}
\bibinfo{author}{\bibfnamefont{A.~E.~C.} \bibnamefont{Hernández}}
  \bibnamefont{and} \bibinfo{author}{\bibfnamefont{I.}~\bibnamefont{Nisandzic}}
  (\bibinfo{year}{2015}), \eprint{arXiv:1512.07165}.

\bibitem[{\citenamefont{Dey et~al.}(2015)\citenamefont{Dey, Mohanty, and
  Tomar}}]{Dey:2015bur}
\bibinfo{author}{\bibfnamefont{U.~K.} \bibnamefont{Dey}},
  \bibinfo{author}{\bibfnamefont{S.}~\bibnamefont{Mohanty}}, \bibnamefont{and}
  \bibinfo{author}{\bibfnamefont{G.}~\bibnamefont{Tomar}}
  (\bibinfo{year}{2015}), \eprint{arXiv:1512.07212}.

\bibitem[{\citenamefont{Pelaggi et~al.}(2015)\citenamefont{Pelaggi, Strumia,
  and Vigiani}}]{Pelaggi:2015knk}
\bibinfo{author}{\bibfnamefont{G.~M.} \bibnamefont{Pelaggi}},
  \bibinfo{author}{\bibfnamefont{A.}~\bibnamefont{Strumia}}, \bibnamefont{and}
  \bibinfo{author}{\bibfnamefont{E.}~\bibnamefont{Vigiani}}
  (\bibinfo{year}{2015}), \eprint{arXiv:1512.07225}.

\bibitem[{\citenamefont{de~Blas et~al.}(2015)\citenamefont{de~Blas, Santiago,
  and Vega-Morales}}]{deBlas:2015hlv}
\bibinfo{author}{\bibfnamefont{J.}~\bibnamefont{de~Blas}},
  \bibinfo{author}{\bibfnamefont{J.}~\bibnamefont{Santiago}}, \bibnamefont{and}
  \bibinfo{author}{\bibfnamefont{R.}~\bibnamefont{Vega-Morales}}
  (\bibinfo{year}{2015}), \eprint{arXiv:1512.07229}.

\bibitem[{\citenamefont{Belyaev et~al.}(2015)\citenamefont{Belyaev,
  Cacciapaglia, Cai, Flacke, Parolini, and Serôdio}}]{Belyaev:2015hgo}
\bibinfo{author}{\bibfnamefont{A.}~\bibnamefont{Belyaev}},
  \bibinfo{author}{\bibfnamefont{G.}~\bibnamefont{Cacciapaglia}},
  \bibinfo{author}{\bibfnamefont{H.}~\bibnamefont{Cai}},
  \bibinfo{author}{\bibfnamefont{T.}~\bibnamefont{Flacke}},
  \bibinfo{author}{\bibfnamefont{A.}~\bibnamefont{Parolini}}, \bibnamefont{and}
  \bibinfo{author}{\bibfnamefont{H.}~\bibnamefont{Serôdio}}
  (\bibinfo{year}{2015}), \eprint{arXiv:1512.07242}.

\bibitem[{\citenamefont{Dev and Teresi}(2015)}]{Dev:2015isx}
\bibinfo{author}{\bibfnamefont{P.~S.~B.} \bibnamefont{Dev}} \bibnamefont{and}
  \bibinfo{author}{\bibfnamefont{D.}~\bibnamefont{Teresi}}
  (\bibinfo{year}{2015}), \eprint{arXiv:1512.07243}.

\bibitem[{\citenamefont{Huang et~al.}(2015{\natexlab{b}})\citenamefont{Huang,
  Tsai, and Yuan}}]{Huang:2015rkj}
\bibinfo{author}{\bibfnamefont{W.-C.} \bibnamefont{Huang}},
  \bibinfo{author}{\bibfnamefont{Y.-L.~S.} \bibnamefont{Tsai}},
  \bibnamefont{and} \bibinfo{author}{\bibfnamefont{T.-C.} \bibnamefont{Yuan}}
  (\bibinfo{year}{2015}{\natexlab{b}}), \eprint{arXiv:1512.07268}.

\bibitem[{\citenamefont{Moretti and Yagyu}(2015)}]{Moretti:2015pbj}
\bibinfo{author}{\bibfnamefont{S.}~\bibnamefont{Moretti}} \bibnamefont{and}
  \bibinfo{author}{\bibfnamefont{K.}~\bibnamefont{Yagyu}}
  (\bibinfo{year}{2015}), \eprint{arXiv:1512.07462}.

\bibitem[{\citenamefont{Patel and Sharma}(2015)}]{Patel:2015ulo}
\bibinfo{author}{\bibfnamefont{K.~M.} \bibnamefont{Patel}} \bibnamefont{and}
  \bibinfo{author}{\bibfnamefont{P.}~\bibnamefont{Sharma}}
  (\bibinfo{year}{2015}), \eprint{arXiv:1512.07468}.

\bibitem[{\citenamefont{Badziak}(2015)}]{Badziak:2015zez}
\bibinfo{author}{\bibfnamefont{M.}~\bibnamefont{Badziak}}
  (\bibinfo{year}{2015}), \eprint{arXiv:1512.07497}.

\bibitem[{\citenamefont{Chakraborty et~al.}(2015)\citenamefont{Chakraborty,
  Chakraborty, and Raychaudhuri}}]{Chakraborty:2015gyj}
\bibinfo{author}{\bibfnamefont{S.}~\bibnamefont{Chakraborty}},
  \bibinfo{author}{\bibfnamefont{A.}~\bibnamefont{Chakraborty}},
  \bibnamefont{and}
  \bibinfo{author}{\bibfnamefont{S.}~\bibnamefont{Raychaudhuri}}
  (\bibinfo{year}{2015}), \eprint{arXiv:1512.07527}.

\bibitem[{\citenamefont{Cao et~al.}(2015{\natexlab{c}})\citenamefont{Cao, Chen,
  and Gu}}]{Cao:2015xjz}
\bibinfo{author}{\bibfnamefont{Q.-H.} \bibnamefont{Cao}},
  \bibinfo{author}{\bibfnamefont{S.-L.} \bibnamefont{Chen}}, \bibnamefont{and}
  \bibinfo{author}{\bibfnamefont{P.-H.} \bibnamefont{Gu}}
  (\bibinfo{year}{2015}{\natexlab{c}}), \eprint{arXiv:1512.07541}.

\bibitem[{\citenamefont{Altmannshofer et~al.}(2015)\citenamefont{Altmannshofer,
  Galloway, Gori, Kagan, Martin, and Zupan}}]{Altmannshofer:2015xfo}
\bibinfo{author}{\bibfnamefont{W.}~\bibnamefont{Altmannshofer}},
  \bibinfo{author}{\bibfnamefont{J.}~\bibnamefont{Galloway}},
  \bibinfo{author}{\bibfnamefont{S.}~\bibnamefont{Gori}},
  \bibinfo{author}{\bibfnamefont{A.~L.} \bibnamefont{Kagan}},
  \bibinfo{author}{\bibfnamefont{A.}~\bibnamefont{Martin}}, \bibnamefont{and}
  \bibinfo{author}{\bibfnamefont{J.}~\bibnamefont{Zupan}}
  (\bibinfo{year}{2015}), \eprint{arXiv:1512.07616}.

\bibitem[{\citenamefont{Cveti$\text{\v{c}}$
  et~al.}(2015)\citenamefont{Cveti$\text{\v{c}}$, Halverson, and
  Langacker}}]{Cvetic:2015vit}
\bibinfo{author}{\bibfnamefont{M.}~\bibnamefont{Cveti$\text{\v{c}}$}},
  \bibinfo{author}{\bibfnamefont{J.}~\bibnamefont{Halverson}},
  \bibnamefont{and} \bibinfo{author}{\bibfnamefont{P.}~\bibnamefont{Langacker}}
  (\bibinfo{year}{2015}), \eprint{arXiv:1512.07622}.

\bibitem[{\citenamefont{Gu and Liu}(2015)}]{Gu:2015lxj}
\bibinfo{author}{\bibfnamefont{J.}~\bibnamefont{Gu}} \bibnamefont{and}
  \bibinfo{author}{\bibfnamefont{Z.}~\bibnamefont{Liu}} (\bibinfo{year}{2015}),
  \eprint{arXiv:1512.07624}.

\bibitem[{\citenamefont{Allanach et~al.}(2015)\citenamefont{Allanach, Dev,
  Renner, and Sakurai}}]{Allanach:2015ixl}
\bibinfo{author}{\bibfnamefont{B.~C.} \bibnamefont{Allanach}},
  \bibinfo{author}{\bibfnamefont{P.~S.~B.} \bibnamefont{Dev}},
  \bibinfo{author}{\bibfnamefont{S.~A.} \bibnamefont{Renner}},
  \bibnamefont{and} \bibinfo{author}{\bibfnamefont{K.}~\bibnamefont{Sakurai}}
  (\bibinfo{year}{2015}), \eprint{arXiv:1512.07645}.

\bibitem[{\citenamefont{Davoudiasl and Zhang}(2015)}]{Davoudiasl:2015cuo}
\bibinfo{author}{\bibfnamefont{H.}~\bibnamefont{Davoudiasl}} \bibnamefont{and}
  \bibinfo{author}{\bibfnamefont{C.}~\bibnamefont{Zhang}}
  (\bibinfo{year}{2015}), \eprint{arXiv:1512.07672}.

\bibitem[{\citenamefont{Craig et~al.}(2015{\natexlab{a}})\citenamefont{Craig,
  Draper, Kilic, and Thomas}}]{Craig:2015lra}
\bibinfo{author}{\bibfnamefont{N.}~\bibnamefont{Craig}},
  \bibinfo{author}{\bibfnamefont{P.}~\bibnamefont{Draper}},
  \bibinfo{author}{\bibfnamefont{C.}~\bibnamefont{Kilic}}, \bibnamefont{and}
  \bibinfo{author}{\bibfnamefont{S.}~\bibnamefont{Thomas}}
  (\bibinfo{year}{2015}{\natexlab{a}}), \eprint{arXiv:1512.07733}.

\bibitem[{\citenamefont{Das and Rai}(2015)}]{Das:2015enc}
\bibinfo{author}{\bibfnamefont{K.}~\bibnamefont{Das}} \bibnamefont{and}
  \bibinfo{author}{\bibfnamefont{S.~K.} \bibnamefont{Rai}}
  (\bibinfo{year}{2015}), \eprint{arXiv:1512.07789}.

\bibitem[{\citenamefont{Cheung et~al.}(2015)\citenamefont{Cheung, Ko, Lee,
  Park, and Tseng}}]{Cheung:2015cug}
\bibinfo{author}{\bibfnamefont{K.}~\bibnamefont{Cheung}},
  \bibinfo{author}{\bibfnamefont{P.}~\bibnamefont{Ko}},
  \bibinfo{author}{\bibfnamefont{J.~S.} \bibnamefont{Lee}},
  \bibinfo{author}{\bibfnamefont{J.}~\bibnamefont{Park}}, \bibnamefont{and}
  \bibinfo{author}{\bibfnamefont{P.-Y.} \bibnamefont{Tseng}}
  (\bibinfo{year}{2015}), \eprint{arXiv:1512.07853}.

\bibitem[{\citenamefont{Liu et~al.}(2015)\citenamefont{Liu, Wang, and
  Xue}}]{Liu:2015yec}
\bibinfo{author}{\bibfnamefont{J.}~\bibnamefont{Liu}},
  \bibinfo{author}{\bibfnamefont{X.-P.} \bibnamefont{Wang}}, \bibnamefont{and}
  \bibinfo{author}{\bibfnamefont{W.}~\bibnamefont{Xue}} (\bibinfo{year}{2015}),
  \eprint{arXiv:1512.07885}.

\bibitem[{\citenamefont{Zhang and Zhou}(2015)}]{Zhang:2015uuo}
\bibinfo{author}{\bibfnamefont{J.}~\bibnamefont{Zhang}} \bibnamefont{and}
  \bibinfo{author}{\bibfnamefont{S.}~\bibnamefont{Zhou}}
  (\bibinfo{year}{2015}), \eprint{arXiv:1512.07889}.

\bibitem[{\citenamefont{Casas et~al.}(2015)\citenamefont{Casas, Espinosa, and
  Moreno}}]{Casas:2015blx}
\bibinfo{author}{\bibfnamefont{J.~A.} \bibnamefont{Casas}},
  \bibinfo{author}{\bibfnamefont{J.~R.} \bibnamefont{Espinosa}},
  \bibnamefont{and} \bibinfo{author}{\bibfnamefont{J.~M.} \bibnamefont{Moreno}}
  (\bibinfo{year}{2015}), \eprint{arXiv:1512.07895}.

\bibitem[{\citenamefont{Hall et~al.}(2015)\citenamefont{Hall, Harigaya, and
  Nomura}}]{Hall:2015xds}
\bibinfo{author}{\bibfnamefont{L.~J.} \bibnamefont{Hall}},
  \bibinfo{author}{\bibfnamefont{K.}~\bibnamefont{Harigaya}}, \bibnamefont{and}
  \bibinfo{author}{\bibfnamefont{Y.}~\bibnamefont{Nomura}}
  (\bibinfo{year}{2015}), \eprint{arXiv:1512.07904}.

\bibitem[{\citenamefont{Han et~al.}(2015{\natexlab{c}})\citenamefont{Han, Wang,
  and Zheng}}]{Han:2015yjk}
\bibinfo{author}{\bibfnamefont{H.}~\bibnamefont{Han}},
  \bibinfo{author}{\bibfnamefont{S.}~\bibnamefont{Wang}}, \bibnamefont{and}
  \bibinfo{author}{\bibfnamefont{S.}~\bibnamefont{Zheng}}
  (\bibinfo{year}{2015}{\natexlab{c}}), \eprint{arXiv:1512.07992}.

\bibitem[{\citenamefont{Park and Park}(2015)}]{Park:2015ysf}
\bibinfo{author}{\bibfnamefont{J.-C.} \bibnamefont{Park}} \bibnamefont{and}
  \bibinfo{author}{\bibfnamefont{S.~C.} \bibnamefont{Park}}
  (\bibinfo{year}{2015}), \eprint{arXiv:1512.08117}.

\bibitem[{\citenamefont{Salvio and Mazumdar}(2015)}]{Salvio:2015jgu}
\bibinfo{author}{\bibfnamefont{A.}~\bibnamefont{Salvio}} \bibnamefont{and}
  \bibinfo{author}{\bibfnamefont{A.}~\bibnamefont{Mazumdar}}
  (\bibinfo{year}{2015}), \eprint{arXiv:1512.08184}.

\bibitem[{\citenamefont{Chway et~al.}(2015)\citenamefont{Chway, Dermí¨ek, Jung,
  and Kim}}]{Chway:2015lzg}
\bibinfo{author}{\bibfnamefont{D.}~\bibnamefont{Chway}},
  \bibinfo{author}{\bibfnamefont{R.}~\bibnamefont{Dermí¨ek}},
  \bibinfo{author}{\bibfnamefont{T.~H.} \bibnamefont{Jung}}, \bibnamefont{and}
  \bibinfo{author}{\bibfnamefont{H.~D.} \bibnamefont{Kim}}
  (\bibinfo{year}{2015}), \eprint{arXiv:1512.08221}.

\bibitem[{\citenamefont{Li et~al.}(2015)\citenamefont{Li, Mao, Tang, Zhang,
  Zhou, and Zhu}}]{Li:2015jwd}
\bibinfo{author}{\bibfnamefont{G.}~\bibnamefont{Li}},
  \bibinfo{author}{\bibfnamefont{Y.-n.} \bibnamefont{Mao}},
  \bibinfo{author}{\bibfnamefont{Y.-L.} \bibnamefont{Tang}},
  \bibinfo{author}{\bibfnamefont{C.}~\bibnamefont{Zhang}},
  \bibinfo{author}{\bibfnamefont{Y.}~\bibnamefont{Zhou}}, \bibnamefont{and}
  \bibinfo{author}{\bibfnamefont{S.-h.} \bibnamefont{Zhu}}
  (\bibinfo{year}{2015}), \eprint{arXiv:1512.08255}.

\bibitem[{\citenamefont{Son and Urbano}(2015)}]{Son:2015vfl}
\bibinfo{author}{\bibfnamefont{M.}~\bibnamefont{Son}} \bibnamefont{and}
  \bibinfo{author}{\bibfnamefont{A.}~\bibnamefont{Urbano}}
  (\bibinfo{year}{2015}), \eprint{arXiv:1512.08307}.

\bibitem[{\citenamefont{Tang and Zhu}(2015)}]{Tang:2015eko}
\bibinfo{author}{\bibfnamefont{Y.-L.} \bibnamefont{Tang}} \bibnamefont{and}
  \bibinfo{author}{\bibfnamefont{S.-h.} \bibnamefont{Zhu}}
  (\bibinfo{year}{2015}), \eprint{arXiv:1512.08323}.

\bibitem[{\citenamefont{An et~al.}(2015)\citenamefont{An, Cheung, and
  Zhang}}]{An:2015cgp}
\bibinfo{author}{\bibfnamefont{H.}~\bibnamefont{An}},
  \bibinfo{author}{\bibfnamefont{C.}~\bibnamefont{Cheung}}, \bibnamefont{and}
  \bibinfo{author}{\bibfnamefont{Y.}~\bibnamefont{Zhang}}
  (\bibinfo{year}{2015}), \eprint{arXiv:1512.08378}.

\bibitem[{\citenamefont{Cao et~al.}(2015{\natexlab{d}})\citenamefont{Cao, Wang,
  and Zhang}}]{Cao:2015apa}
\bibinfo{author}{\bibfnamefont{J.}~\bibnamefont{Cao}},
  \bibinfo{author}{\bibfnamefont{F.}~\bibnamefont{Wang}}, \bibnamefont{and}
  \bibinfo{author}{\bibfnamefont{Y.}~\bibnamefont{Zhang}}
  (\bibinfo{year}{2015}{\natexlab{d}}), \eprint{arXiv:1512.08392}.

\bibitem[{\citenamefont{Wang et~al.}(2015{\natexlab{b}})\citenamefont{Wang,
  Wang, Wu, Yang, and Zhang}}]{Wang:2015omi}
\bibinfo{author}{\bibfnamefont{F.}~\bibnamefont{Wang}},
  \bibinfo{author}{\bibfnamefont{W.}~\bibnamefont{Wang}},
  \bibinfo{author}{\bibfnamefont{L.}~\bibnamefont{Wu}},
  \bibinfo{author}{\bibfnamefont{J.~M.} \bibnamefont{Yang}}, \bibnamefont{and}
  \bibinfo{author}{\bibfnamefont{M.}~\bibnamefont{Zhang}}
  (\bibinfo{year}{2015}{\natexlab{b}}), \eprint{arXiv:1512.08434}.

\bibitem[{\citenamefont{Cai et~al.}(2015)\citenamefont{Cai, Yu, and
  Zhang}}]{Cai:2015hzc}
\bibinfo{author}{\bibfnamefont{C.}~\bibnamefont{Cai}},
  \bibinfo{author}{\bibfnamefont{Z.-H.} \bibnamefont{Yu}}, \bibnamefont{and}
  \bibinfo{author}{\bibfnamefont{H.-H.} \bibnamefont{Zhang}}
  (\bibinfo{year}{2015}), \eprint{arXiv:1512.08440}.

\bibitem[{\citenamefont{Cao et~al.}(2015{\natexlab{e}})\citenamefont{Cao, Liu,
  Xie, Yan, and Zhang}}]{Cao:2015scs}
\bibinfo{author}{\bibfnamefont{Q.-H.} \bibnamefont{Cao}},
  \bibinfo{author}{\bibfnamefont{Y.}~\bibnamefont{Liu}},
  \bibinfo{author}{\bibfnamefont{K.-P.} \bibnamefont{Xie}},
  \bibinfo{author}{\bibfnamefont{B.}~\bibnamefont{Yan}}, \bibnamefont{and}
  \bibinfo{author}{\bibfnamefont{D.-M.} \bibnamefont{Zhang}}
  (\bibinfo{year}{2015}{\natexlab{e}}), \eprint{arXiv:1512.08441}.

\bibitem[{\citenamefont{Kim}(2015)}]{Kim:2015xyn}
\bibinfo{author}{\bibfnamefont{J.~E.} \bibnamefont{Kim}}
  (\bibinfo{year}{2015}), \eprint{arXiv:1512.08467}.

\bibitem[{\citenamefont{Gao et~al.}(2015)\citenamefont{Gao, Zhang, and
  Zhu}}]{Gao:2015igz}
\bibinfo{author}{\bibfnamefont{J.}~\bibnamefont{Gao}},
  \bibinfo{author}{\bibfnamefont{H.}~\bibnamefont{Zhang}}, \bibnamefont{and}
  \bibinfo{author}{\bibfnamefont{H.~X.} \bibnamefont{Zhu}}
  (\bibinfo{year}{2015}), \eprint{arXiv:1512.08478}.

\bibitem[{\citenamefont{Chao}(2015{\natexlab{b}})}]{Chao:2015nac}
\bibinfo{author}{\bibfnamefont{W.}~\bibnamefont{Chao}}
  (\bibinfo{year}{2015}{\natexlab{b}}), \eprint{arXiv:1512.08484}.

\bibitem[{\citenamefont{Bi et~al.}(2015{\natexlab{b}})\citenamefont{Bi, Ding,
  Fan, Huang, Li, Li, Raza, Wang, and Zhu}}]{Bi:2015lcf}
\bibinfo{author}{\bibfnamefont{X.-J.} \bibnamefont{Bi}},
  \bibinfo{author}{\bibfnamefont{R.}~\bibnamefont{Ding}},
  \bibinfo{author}{\bibfnamefont{Y.}~\bibnamefont{Fan}},
  \bibinfo{author}{\bibfnamefont{L.}~\bibnamefont{Huang}},
  \bibinfo{author}{\bibfnamefont{C.}~\bibnamefont{Li}},
  \bibinfo{author}{\bibfnamefont{T.}~\bibnamefont{Li}},
  \bibinfo{author}{\bibfnamefont{S.}~\bibnamefont{Raza}},
  \bibinfo{author}{\bibfnamefont{X.-C.} \bibnamefont{Wang}}, \bibnamefont{and}
  \bibinfo{author}{\bibfnamefont{B.}~\bibnamefont{Zhu}}
  (\bibinfo{year}{2015}{\natexlab{b}}), \eprint{arXiv:1512.08497}.

\bibitem[{\citenamefont{Goertz et~al.}(2015)\citenamefont{Goertz, Kamenik,
  Katz, and Nardecchia}}]{Goertz:2015nkp}
\bibinfo{author}{\bibfnamefont{F.}~\bibnamefont{Goertz}},
  \bibinfo{author}{\bibfnamefont{J.~F.} \bibnamefont{Kamenik}},
  \bibinfo{author}{\bibfnamefont{A.}~\bibnamefont{Katz}}, \bibnamefont{and}
  \bibinfo{author}{\bibfnamefont{M.}~\bibnamefont{Nardecchia}}
  (\bibinfo{year}{2015}), \eprint{arXiv:1512.08500}.

\bibitem[{\citenamefont{Anchordoqui et~al.}(2015)\citenamefont{Anchordoqui,
  Antoniadis, Goldberg, Huang, Lust, and Taylor}}]{Anchordoqui:2015jxc}
\bibinfo{author}{\bibfnamefont{L.~A.} \bibnamefont{Anchordoqui}},
  \bibinfo{author}{\bibfnamefont{I.}~\bibnamefont{Antoniadis}},
  \bibinfo{author}{\bibfnamefont{H.}~\bibnamefont{Goldberg}},
  \bibinfo{author}{\bibfnamefont{X.}~\bibnamefont{Huang}},
  \bibinfo{author}{\bibfnamefont{D.}~\bibnamefont{Lust}}, \bibnamefont{and}
  \bibinfo{author}{\bibfnamefont{T.~R.} \bibnamefont{Taylor}}
  (\bibinfo{year}{2015}), \eprint{arXiv:1512.08502}.

\bibitem[{\citenamefont{Dev et~al.}(2015)\citenamefont{Dev, Mohapatra, and
  Zhang}}]{Dev:2015vjd}
\bibinfo{author}{\bibfnamefont{P.~S.~B.} \bibnamefont{Dev}},
  \bibinfo{author}{\bibfnamefont{R.~N.} \bibnamefont{Mohapatra}},
  \bibnamefont{and} \bibinfo{author}{\bibfnamefont{Y.}~\bibnamefont{Zhang}}
  (\bibinfo{year}{2015}), \eprint{arXiv:1512.08507}.

\bibitem[{\citenamefont{Bizot et~al.}(2015)\citenamefont{Bizot, Davidson,
  Frigerio, and Kneur}}]{Bizot:2015qqo}
\bibinfo{author}{\bibfnamefont{N.}~\bibnamefont{Bizot}},
  \bibinfo{author}{\bibfnamefont{S.}~\bibnamefont{Davidson}},
  \bibinfo{author}{\bibfnamefont{M.}~\bibnamefont{Frigerio}}, \bibnamefont{and}
  \bibinfo{author}{\bibfnamefont{J.~L.} \bibnamefont{Kneur}}
  (\bibinfo{year}{2015}), \eprint{arXiv:1512.08508}.

\bibitem[{\citenamefont{Ibanez and Martin-Lozano}(2015)}]{Ibanez:2015uok}
\bibinfo{author}{\bibfnamefont{L.~E.} \bibnamefont{Ibanez}} \bibnamefont{and}
  \bibinfo{author}{\bibfnamefont{V.}~\bibnamefont{Martin-Lozano}}
  (\bibinfo{year}{2015}), \eprint{arXiv:1512.08777}.

\bibitem[{\citenamefont{Chiang et~al.}(2015)\citenamefont{Chiang, Ibe, and
  Yanagida}}]{Chiang:2015tqz}
\bibinfo{author}{\bibfnamefont{C.-W.} \bibnamefont{Chiang}},
  \bibinfo{author}{\bibfnamefont{M.}~\bibnamefont{Ibe}}, \bibnamefont{and}
  \bibinfo{author}{\bibfnamefont{T.~T.} \bibnamefont{Yanagida}}
  (\bibinfo{year}{2015}), \eprint{arXiv:1512.08895}.

\bibitem[{\citenamefont{Kang and Song}(2015)}]{Kang:2015roj}
\bibinfo{author}{\bibfnamefont{S.~K.} \bibnamefont{Kang}} \bibnamefont{and}
  \bibinfo{author}{\bibfnamefont{J.}~\bibnamefont{Song}}
  (\bibinfo{year}{2015}), \eprint{arXiv:1512.08963}.

\bibitem[{\citenamefont{Hamada et~al.}(2015)\citenamefont{Hamada, Noumi, Sun,
  and Shiu}}]{Hamada:2015skp}
\bibinfo{author}{\bibfnamefont{Y.}~\bibnamefont{Hamada}},
  \bibinfo{author}{\bibfnamefont{T.}~\bibnamefont{Noumi}},
  \bibinfo{author}{\bibfnamefont{S.}~\bibnamefont{Sun}}, \bibnamefont{and}
  \bibinfo{author}{\bibfnamefont{G.}~\bibnamefont{Shiu}}
  (\bibinfo{year}{2015}), \eprint{arXiv:1512.08984}.

\bibitem[{\citenamefont{Huang et~al.}(2015{\natexlab{c}})\citenamefont{Huang,
  Zhang, and Zhou}}]{Huang:2015svl}
\bibinfo{author}{\bibfnamefont{X.-J.} \bibnamefont{Huang}},
  \bibinfo{author}{\bibfnamefont{W.-H.} \bibnamefont{Zhang}}, \bibnamefont{and}
  \bibinfo{author}{\bibfnamefont{Y.-F.} \bibnamefont{Zhou}}
  (\bibinfo{year}{2015}{\natexlab{c}}), \eprint{arXiv:1512.08992}.

\bibitem[{\citenamefont{Kanemura
  et~al.}(2015{\natexlab{a}})\citenamefont{Kanemura, Nishiwaki, Okada, Orikasa,
  Park, and Watanabe}}]{Kanemura:2015bli}
\bibinfo{author}{\bibfnamefont{S.}~\bibnamefont{Kanemura}},
  \bibinfo{author}{\bibfnamefont{K.}~\bibnamefont{Nishiwaki}},
  \bibinfo{author}{\bibfnamefont{H.}~\bibnamefont{Okada}},
  \bibinfo{author}{\bibfnamefont{Y.}~\bibnamefont{Orikasa}},
  \bibinfo{author}{\bibfnamefont{S.~C.} \bibnamefont{Park}}, \bibnamefont{and}
  \bibinfo{author}{\bibfnamefont{R.}~\bibnamefont{Watanabe}}
  (\bibinfo{year}{2015}{\natexlab{a}}), \eprint{arXiv:1512.09048}.

\bibitem[{\citenamefont{Kanemura
  et~al.}(2015{\natexlab{b}})\citenamefont{Kanemura, Machida, Odori, and
  Shindou}}]{Kanemura:2015vcb}
\bibinfo{author}{\bibfnamefont{S.}~\bibnamefont{Kanemura}},
  \bibinfo{author}{\bibfnamefont{N.}~\bibnamefont{Machida}},
  \bibinfo{author}{\bibfnamefont{S.}~\bibnamefont{Odori}}, \bibnamefont{and}
  \bibinfo{author}{\bibfnamefont{T.}~\bibnamefont{Shindou}}
  (\bibinfo{year}{2015}{\natexlab{b}}), \eprint{arXiv:1512.09053}.

\bibitem[{\citenamefont{Low and Lykken}(2015)}]{Low:2015qho}
\bibinfo{author}{\bibfnamefont{I.}~\bibnamefont{Low}} \bibnamefont{and}
  \bibinfo{author}{\bibfnamefont{J.}~\bibnamefont{Lykken}}
  (\bibinfo{year}{2015}), \eprint{arXiv:1512.09089}.

\bibitem[{\citenamefont{Hernández}(2015)}]{Hernandez:2015hrt}
\bibinfo{author}{\bibfnamefont{A.~E.~C.} \bibnamefont{Hernández}}
  (\bibinfo{year}{2015}), \eprint{arXiv:1512.09092}.

\bibitem[{\citenamefont{Jiang et~al.}(2015)\citenamefont{Jiang, Li, and
  Liu}}]{Jiang:2015oms}
\bibinfo{author}{\bibfnamefont{Y.}~\bibnamefont{Jiang}},
  \bibinfo{author}{\bibfnamefont{Y.-Y.} \bibnamefont{Li}}, \bibnamefont{and}
  \bibinfo{author}{\bibfnamefont{T.}~\bibnamefont{Liu}} (\bibinfo{year}{2015}),
  \eprint{arXiv:1512.09127}.

\bibitem[{\citenamefont{Kaneta et~al.}(2015)\citenamefont{Kaneta, Kang, and
  Lee}}]{Kaneta:2015qpf}
\bibinfo{author}{\bibfnamefont{K.}~\bibnamefont{Kaneta}},
  \bibinfo{author}{\bibfnamefont{S.}~\bibnamefont{Kang}}, \bibnamefont{and}
  \bibinfo{author}{\bibfnamefont{H.-S.} \bibnamefont{Lee}}
  (\bibinfo{year}{2015}), \eprint{arXiv:1512.09129}.

\bibitem[{\citenamefont{Marzola et~al.}(2015)\citenamefont{Marzola, Racioppi,
  Raidal, Urban, and Veermäe}}]{Marzola:2015xbh}
\bibinfo{author}{\bibfnamefont{L.}~\bibnamefont{Marzola}},
  \bibinfo{author}{\bibfnamefont{A.}~\bibnamefont{Racioppi}},
  \bibinfo{author}{\bibfnamefont{M.}~\bibnamefont{Raidal}},
  \bibinfo{author}{\bibfnamefont{F.~R.} \bibnamefont{Urban}}, \bibnamefont{and}
  \bibinfo{author}{\bibfnamefont{H.}~\bibnamefont{Veermäe}}
  (\bibinfo{year}{2015}), \eprint{arXiv:1512.09136}.

\bibitem[{\citenamefont{Ma}(2015)}]{Ma:2015xmf}
\bibinfo{author}{\bibfnamefont{E.}~\bibnamefont{Ma}} (\bibinfo{year}{2015}),
  \eprint{arXiv:1512.09159}.

\bibitem[{\citenamefont{Dasgupta et~al.}(2015)\citenamefont{Dasgupta, Mitra,
  and Borah}}]{Dasgupta:2015pbr}
\bibinfo{author}{\bibfnamefont{A.}~\bibnamefont{Dasgupta}},
  \bibinfo{author}{\bibfnamefont{M.}~\bibnamefont{Mitra}}, \bibnamefont{and}
  \bibinfo{author}{\bibfnamefont{D.}~\bibnamefont{Borah}}
  (\bibinfo{year}{2015}), \eprint{arXiv:1512.09202}.

\bibitem[{\citenamefont{Jung et~al.}(2015)\citenamefont{Jung, Song, and
  Yoon}}]{Jung:2015etr}
\bibinfo{author}{\bibfnamefont{S.}~\bibnamefont{Jung}},
  \bibinfo{author}{\bibfnamefont{J.}~\bibnamefont{Song}}, \bibnamefont{and}
  \bibinfo{author}{\bibfnamefont{Y.~W.} \bibnamefont{Yoon}}
  (\bibinfo{year}{2015}), \eprint{1601.00006}.

\bibitem[{\citenamefont{Potter}(2016)}]{Potter:2016psi}
\bibinfo{author}{\bibfnamefont{C.~T.} \bibnamefont{Potter}}
  (\bibinfo{year}{2016}), \eprint{1601.00240}.

\bibitem[{\citenamefont{Palti}(2016)}]{Palti:2016kew}
\bibinfo{author}{\bibfnamefont{E.}~\bibnamefont{Palti}} (\bibinfo{year}{2016}),
  \eprint{1601.00285}.

\bibitem[{\citenamefont{Nomura and Okada}(2016)}]{Nomura:2016fzs}
\bibinfo{author}{\bibfnamefont{T.}~\bibnamefont{Nomura}} \bibnamefont{and}
  \bibinfo{author}{\bibfnamefont{H.}~\bibnamefont{Okada}}
  (\bibinfo{year}{2016}), \eprint{1601.00386}.

\bibitem[{\citenamefont{Han et~al.}(2016)\citenamefont{Han, Wang, Wu, Yang, and
  Zhang}}]{Han:2016bus}
\bibinfo{author}{\bibfnamefont{X.-F.} \bibnamefont{Han}},
  \bibinfo{author}{\bibfnamefont{L.}~\bibnamefont{Wang}},
  \bibinfo{author}{\bibfnamefont{L.}~\bibnamefont{Wu}},
  \bibinfo{author}{\bibfnamefont{J.~M.} \bibnamefont{Yang}}, \bibnamefont{and}
  \bibinfo{author}{\bibfnamefont{M.}~\bibnamefont{Zhang}}
  (\bibinfo{year}{2016}), \eprint{1601.00534}.

\bibitem[{\citenamefont{Ko et~al.}(2016)\citenamefont{Ko, Omura, and
  Yu}}]{Ko:2016lai}
\bibinfo{author}{\bibfnamefont{P.}~\bibnamefont{Ko}},
  \bibinfo{author}{\bibfnamefont{Y.}~\bibnamefont{Omura}}, \bibnamefont{and}
  \bibinfo{author}{\bibfnamefont{C.}~\bibnamefont{Yu}} (\bibinfo{year}{2016}),
  \eprint{1601.00586}.

\bibitem[{\citenamefont{Ghorbani and Ghorbani}(2016)}]{Ghorbani:2016jdq}
\bibinfo{author}{\bibfnamefont{K.}~\bibnamefont{Ghorbani}} \bibnamefont{and}
  \bibinfo{author}{\bibfnamefont{H.}~\bibnamefont{Ghorbani}}
  (\bibinfo{year}{2016}), \eprint{1601.00602}.

\bibitem[{\citenamefont{Danielsson et~al.}(2016)\citenamefont{Danielsson,
  Enberg, Ingelman, and Mandal}}]{Danielsson:2016nyy}
\bibinfo{author}{\bibfnamefont{U.}~\bibnamefont{Danielsson}},
  \bibinfo{author}{\bibfnamefont{R.}~\bibnamefont{Enberg}},
  \bibinfo{author}{\bibfnamefont{G.}~\bibnamefont{Ingelman}}, \bibnamefont{and}
  \bibinfo{author}{\bibfnamefont{T.}~\bibnamefont{Mandal}}
  (\bibinfo{year}{2016}), \eprint{1601.00624}.

\bibitem[{\citenamefont{Chao}(2016)}]{Chao:2016mtn}
\bibinfo{author}{\bibfnamefont{W.}~\bibnamefont{Chao}} (\bibinfo{year}{2016}),
  \eprint{1601.00633}.

\bibitem[{\citenamefont{Csaki et~al.}(2016)\citenamefont{Csaki, Hubisz,
  Lombardo, and Terning}}]{Csaki:2016raa}
\bibinfo{author}{\bibfnamefont{C.}~\bibnamefont{Csaki}},
  \bibinfo{author}{\bibfnamefont{J.}~\bibnamefont{Hubisz}},
  \bibinfo{author}{\bibfnamefont{S.}~\bibnamefont{Lombardo}}, \bibnamefont{and}
  \bibinfo{author}{\bibfnamefont{J.}~\bibnamefont{Terning}}
  (\bibinfo{year}{2016}), \eprint{1601.00638}.

\bibitem[{\citenamefont{Karozas et~al.}(2016)\citenamefont{Karozas, King,
  Leontaris, and Meadowcroft}}]{Karozas:2016hcp}
\bibinfo{author}{\bibfnamefont{A.}~\bibnamefont{Karozas}},
  \bibinfo{author}{\bibfnamefont{S.~F.} \bibnamefont{King}},
  \bibinfo{author}{\bibfnamefont{G.~K.} \bibnamefont{Leontaris}},
  \bibnamefont{and} \bibinfo{author}{\bibfnamefont{A.~K.}
  \bibnamefont{Meadowcroft}} (\bibinfo{year}{2016}), \eprint{1601.00640}.

\bibitem[{\citenamefont{Hernández et~al.}(2016)\citenamefont{Hernández,
  Varzielas, and Schumacher}}]{Hernandez:2016rbi}
\bibinfo{author}{\bibfnamefont{A.~E.~C.} \bibnamefont{Hernández}},
  \bibinfo{author}{\bibfnamefont{I.~d.~M.} \bibnamefont{Varzielas}},
  \bibnamefont{and}
  \bibinfo{author}{\bibfnamefont{E.}~\bibnamefont{Schumacher}}
  (\bibinfo{year}{2016}), \eprint{1601.00661}.

\bibitem[{\citenamefont{Modak et~al.}(2016)\citenamefont{Modak, Sadhukhan, and
  Srivastava}}]{Modak:2016ung}
\bibinfo{author}{\bibfnamefont{T.}~\bibnamefont{Modak}},
  \bibinfo{author}{\bibfnamefont{S.}~\bibnamefont{Sadhukhan}},
  \bibnamefont{and}
  \bibinfo{author}{\bibfnamefont{R.}~\bibnamefont{Srivastava}}
  (\bibinfo{year}{2016}), \eprint{1601.00836}.

\bibitem[{\citenamefont{Dutta et~al.}(2016)\citenamefont{Dutta, Gao, Ghosh,
  Gogoladze, Li, Shafi, and Walker}}]{Dutta:2016jqn}
\bibinfo{author}{\bibfnamefont{B.}~\bibnamefont{Dutta}},
  \bibinfo{author}{\bibfnamefont{Y.}~\bibnamefont{Gao}},
  \bibinfo{author}{\bibfnamefont{T.}~\bibnamefont{Ghosh}},
  \bibinfo{author}{\bibfnamefont{I.}~\bibnamefont{Gogoladze}},
  \bibinfo{author}{\bibfnamefont{T.}~\bibnamefont{Li}},
  \bibinfo{author}{\bibfnamefont{Q.}~\bibnamefont{Shafi}}, \bibnamefont{and}
  \bibinfo{author}{\bibfnamefont{J.~W.} \bibnamefont{Walker}}
  (\bibinfo{year}{2016}), \eprint{1601.00866}.

\bibitem[{\citenamefont{Deppisch et~al.}(2016)\citenamefont{Deppisch, Hati,
  Patra, Pritimita, and Sarkar}}]{Deppisch:2016scs}
\bibinfo{author}{\bibfnamefont{F.~F.} \bibnamefont{Deppisch}},
  \bibinfo{author}{\bibfnamefont{C.}~\bibnamefont{Hati}},
  \bibinfo{author}{\bibfnamefont{S.}~\bibnamefont{Patra}},
  \bibinfo{author}{\bibfnamefont{P.}~\bibnamefont{Pritimita}},
  \bibnamefont{and} \bibinfo{author}{\bibfnamefont{U.}~\bibnamefont{Sarkar}}
  (\bibinfo{year}{2016}), \eprint{1601.00952}.

\bibitem[{\citenamefont{Craig et~al.}(2015{\natexlab{b}})\citenamefont{Craig,
  D'Eramo, Draper, Thomas, and Zhang}}]{Craig:2015jba}
\bibinfo{author}{\bibfnamefont{N.}~\bibnamefont{Craig}},
  \bibinfo{author}{\bibfnamefont{F.}~\bibnamefont{D'Eramo}},
  \bibinfo{author}{\bibfnamefont{P.}~\bibnamefont{Draper}},
  \bibinfo{author}{\bibfnamefont{S.}~\bibnamefont{Thomas}}, \bibnamefont{and}
  \bibinfo{author}{\bibfnamefont{H.}~\bibnamefont{Zhang}},
  \bibinfo{journal}{JHEP} \textbf{\bibinfo{volume}{06}}, \bibinfo{pages}{137}
  (\bibinfo{year}{2015}{\natexlab{b}}), \eprint{arXiv:1504.04630}.

\bibitem[{\citenamefont{Bernreuther et~al.}(2015)\citenamefont{Bernreuther,
  Galler, Mellein, Si, and Uwer}}]{Bernreuther:2015fts}
\bibinfo{author}{\bibfnamefont{W.}~\bibnamefont{Bernreuther}},
  \bibinfo{author}{\bibfnamefont{P.}~\bibnamefont{Galler}},
  \bibinfo{author}{\bibfnamefont{C.}~\bibnamefont{Mellein}},
  \bibinfo{author}{\bibfnamefont{Z.~G.} \bibnamefont{Si}}, \bibnamefont{and}
  \bibinfo{author}{\bibfnamefont{P.}~\bibnamefont{Uwer}}
  (\bibinfo{year}{2015}), \eprint{arXiv:1511.05584}.

\bibitem[{\citenamefont{Han et~al.}(2004)\citenamefont{Han, Valencia, and
  Wang}}]{Han:2004zh}
\bibinfo{author}{\bibfnamefont{T.}~\bibnamefont{Han}},
  \bibinfo{author}{\bibfnamefont{G.}~\bibnamefont{Valencia}}, \bibnamefont{and}
  \bibinfo{author}{\bibfnamefont{Y.}~\bibnamefont{Wang}},
  \bibinfo{journal}{Phys. Rev.} \textbf{\bibinfo{volume}{D70}},
  \bibinfo{pages}{034002} (\bibinfo{year}{2004}), \eprint{hep-ph/0405055}.

\bibitem[{\citenamefont{Alwall et~al.}(2014)\citenamefont{Alwall, Frederix,
  Frixione, Hirschi, Maltoni et~al.}}]{Alwall:2014hca}
\bibinfo{author}{\bibfnamefont{J.}~\bibnamefont{Alwall}},
  \bibinfo{author}{\bibfnamefont{R.}~\bibnamefont{Frederix}},
  \bibinfo{author}{\bibfnamefont{S.}~\bibnamefont{Frixione}},
  \bibinfo{author}{\bibfnamefont{V.}~\bibnamefont{Hirschi}},
  \bibinfo{author}{\bibfnamefont{F.}~\bibnamefont{Maltoni}},
  \bibnamefont{et~al.}, \bibinfo{journal}{JHEP}
  \textbf{\bibinfo{volume}{1407}}, \bibinfo{pages}{079} (\bibinfo{year}{2014}),
  \eprint{arXiv:1405.0301}.

\bibitem[{\citenamefont{Dulat et~al.}(2015)\citenamefont{Dulat, Hou, Gao,
  Guzzi, Huston, Nadolsky, Pumplin, Schmidt, Stump, and Yuan}}]{Dulat:2015mca}
\bibinfo{author}{\bibfnamefont{S.}~\bibnamefont{Dulat}},
  \bibinfo{author}{\bibfnamefont{T.~J.} \bibnamefont{Hou}},
  \bibinfo{author}{\bibfnamefont{J.}~\bibnamefont{Gao}},
  \bibinfo{author}{\bibfnamefont{M.}~\bibnamefont{Guzzi}},
  \bibinfo{author}{\bibfnamefont{J.}~\bibnamefont{Huston}},
  \bibinfo{author}{\bibfnamefont{P.}~\bibnamefont{Nadolsky}},
  \bibinfo{author}{\bibfnamefont{J.}~\bibnamefont{Pumplin}},
  \bibinfo{author}{\bibfnamefont{C.}~\bibnamefont{Schmidt}},
  \bibinfo{author}{\bibfnamefont{D.}~\bibnamefont{Stump}}, \bibnamefont{and}
  \bibinfo{author}{\bibfnamefont{C.~P.} \bibnamefont{Yuan}}
  (\bibinfo{year}{2015}), \eprint{arXiv:1506.07443}.

\bibitem[{\citenamefont{Sjostrand et~al.}(2006)\citenamefont{Sjostrand, Mrenna,
  and Skands}}]{Sjostrand:2006za}
\bibinfo{author}{\bibfnamefont{T.}~\bibnamefont{Sjostrand}},
  \bibinfo{author}{\bibfnamefont{S.}~\bibnamefont{Mrenna}}, \bibnamefont{and}
  \bibinfo{author}{\bibfnamefont{P.~Z.} \bibnamefont{Skands}},
  \bibinfo{journal}{JHEP} \textbf{\bibinfo{volume}{0605}}, \bibinfo{pages}{026}
  (\bibinfo{year}{2006}), \eprint{hep-ph/0603175}.

\bibitem[{\citenamefont{Field}(2011)}]{Field:2011iq}
\bibinfo{author}{\bibfnamefont{R.}~\bibnamefont{Field}}, \bibinfo{journal}{Acta
  Phys.Polon.} \textbf{\bibinfo{volume}{B42}}, \bibinfo{pages}{2631}
  (\bibinfo{year}{2011}), \eprint{arXiv:1110.5530}.

\bibitem[{\citenamefont{de~Favereau et~al.}(2014)}]{deFavereau:2013fsa}
\bibinfo{author}{\bibfnamefont{J.}~\bibnamefont{de~Favereau}}
  \bibnamefont{et~al.} (\bibinfo{collaboration}{DELPHES 3}),
  \bibinfo{journal}{JHEP} \textbf{\bibinfo{volume}{1402}}, \bibinfo{pages}{057}
  (\bibinfo{year}{2014}), \eprint{arXiv:1307.6346}.

\bibitem[{\citenamefont{Cacciari et~al.}(2012)\citenamefont{Cacciari, Salam,
  and Soyez}}]{Cacciari:2011ma}
\bibinfo{author}{\bibfnamefont{M.}~\bibnamefont{Cacciari}},
  \bibinfo{author}{\bibfnamefont{G.~P.} \bibnamefont{Salam}}, \bibnamefont{and}
  \bibinfo{author}{\bibfnamefont{G.}~\bibnamefont{Soyez}},
  \bibinfo{journal}{Eur.Phys.J.} \textbf{\bibinfo{volume}{C72}},
  \bibinfo{pages}{1896} (\bibinfo{year}{2012}), \eprint{arXiv:1111.6097}.

\bibitem[{ATL(2015{\natexlab{b}})}]{ATL-PHYS-PUB-2015-022}
\bibinfo{type}{Tech. Rep.} \bibinfo{number}{ATL-PHYS-PUB-2015-022},
  \bibinfo{institution}{CERN}, \bibinfo{address}{Geneva}
  (\bibinfo{year}{2015}{\natexlab{b}}),
  \urlprefix\url{http://cds.cern.ch/record/2037697}.

\bibitem[{\citenamefont{Campbell and Ellis}(2010)}]{Campbell:2010ff}
\bibinfo{author}{\bibfnamefont{J.~M.} \bibnamefont{Campbell}} \bibnamefont{and}
  \bibinfo{author}{\bibfnamefont{R.~K.} \bibnamefont{Ellis}},
  \bibinfo{journal}{Nucl. Phys. Proc. Suppl.}
  \textbf{\bibinfo{volume}{205-206}}, \bibinfo{pages}{10}
  (\bibinfo{year}{2010}), \eprint{arXiv:1007.3492}.

\bibitem[{\citenamefont{Lai et~al.}(2010)\citenamefont{Lai, Guzzi, Huston, Li,
  Nadolsky, Pumplin, and Yuan}}]{Lai:2010vv}
\bibinfo{author}{\bibfnamefont{H.-L.} \bibnamefont{Lai}},
  \bibinfo{author}{\bibfnamefont{M.}~\bibnamefont{Guzzi}},
  \bibinfo{author}{\bibfnamefont{J.}~\bibnamefont{Huston}},
  \bibinfo{author}{\bibfnamefont{Z.}~\bibnamefont{Li}},
  \bibinfo{author}{\bibfnamefont{P.~M.} \bibnamefont{Nadolsky}},
  \bibinfo{author}{\bibfnamefont{J.}~\bibnamefont{Pumplin}}, \bibnamefont{and}
  \bibinfo{author}{\bibfnamefont{C.~P.} \bibnamefont{Yuan}},
  \bibinfo{journal}{Phys. Rev.} \textbf{\bibinfo{volume}{D82}},
  \bibinfo{pages}{074024} (\bibinfo{year}{2010}), \eprint{arXiv:1007.2241}.

\bibitem[{\citenamefont{Martin et~al.}(2009)\citenamefont{Martin, Stirling,
  Thorne, and Watt}}]{Martin:2009iq}
\bibinfo{author}{\bibfnamefont{A.~D.} \bibnamefont{Martin}},
  \bibinfo{author}{\bibfnamefont{W.~J.} \bibnamefont{Stirling}},
  \bibinfo{author}{\bibfnamefont{R.~S.} \bibnamefont{Thorne}},
  \bibnamefont{and} \bibinfo{author}{\bibfnamefont{G.}~\bibnamefont{Watt}},
  \bibinfo{journal}{Eur. Phys. J.} \textbf{\bibinfo{volume}{C63}},
  \bibinfo{pages}{189} (\bibinfo{year}{2009}), \eprint{0901.0002}.

\bibitem[{\citenamefont{Dittmaier et~al.}(2011)}]{Dittmaier:2011ti}
\bibinfo{author}{\bibfnamefont{S.}~\bibnamefont{Dittmaier}}
  \bibnamefont{et~al.} (\bibinfo{collaboration}{LHC Higgs Cross Section Working
  Group}) (\bibinfo{year}{2011}), \eprint{arXiv:1101.0593}.

\bibitem[{ATL(2015{\natexlab{c}})}]{ATLAS-CONF-2015-049}
\bibinfo{type}{Tech. Rep.} \bibinfo{number}{ATLAS-CONF-2015-049},
  \bibinfo{institution}{CERN}, \bibinfo{address}{Geneva}
  (\bibinfo{year}{2015}{\natexlab{c}}),
  \urlprefix\url{http://cds.cern.ch/record/2052605}.

\bibitem[{ATL(2015{\natexlab{d}})}]{ATLAS-CONF-2015-065}
\bibinfo{type}{Tech. Rep.} \bibinfo{number}{ATLAS-CONF-2015-065},
  \bibinfo{institution}{CERN}, \bibinfo{address}{Geneva}
  (\bibinfo{year}{2015}{\natexlab{d}}),
  \urlprefix\url{http://cds.cern.ch/record/2114832}.

\bibitem[{\citenamefont{Read}(2002)}]{Read:2002hq}
\bibinfo{author}{\bibfnamefont{A.~L.} \bibnamefont{Read}}, \bibinfo{journal}{J.
  Phys.} \textbf{\bibinfo{volume}{G28}}, \bibinfo{pages}{2693}
  (\bibinfo{year}{2002}).

\bibitem[{\citenamefont{Cowan et~al.}(2011)\citenamefont{Cowan, Cranmer, Gross,
  and Vitells}}]{Cowan:2010js}
\bibinfo{author}{\bibfnamefont{G.}~\bibnamefont{Cowan}},
  \bibinfo{author}{\bibfnamefont{K.}~\bibnamefont{Cranmer}},
  \bibinfo{author}{\bibfnamefont{E.}~\bibnamefont{Gross}}, \bibnamefont{and}
  \bibinfo{author}{\bibfnamefont{O.}~\bibnamefont{Vitells}},
  \bibinfo{journal}{Eur.Phys.J.} \textbf{\bibinfo{volume}{C71}},
  \bibinfo{pages}{1554} (\bibinfo{year}{2011}), \eprint{arXiv:1007.1727}.

\bibitem[{\citenamefont{Group}(2015{\natexlab{a}})}]{CEPC-SPPCStudyGroup:2015csa}
\bibinfo{author}{\bibfnamefont{CEPC-SPPC Study Group} \bibnamefont{Group}}
  (\bibinfo{year}{2015}{\natexlab{a}}),  \urlprefix\url{http://cepc.ihep.ac.cn/preCDR/main_preCDR.pdf}.

\bibitem[{\citenamefont{Group}(2015{\natexlab{b}})}]{CEPC-SPPCStudyGroup:2015esa}
\bibinfo{author}{\bibfnamefont{CEPC-SPPC Study Group} \bibnamefont{Group}}
  (\bibinfo{year}{2015}{\natexlab{b}}),  \urlprefix\url{http://cepc.ihep.ac.cn/preCDR/Pre-CDR_final_20150317.pdf}.

\end{thebibliography}
  
\end{document}